\documentclass[fleqn,usenatbib]{mnras}
\usepackage{amsmath}
\usepackage{newtxtext,newtxmath}
\usepackage[T1]{fontenc}
\usepackage{ae,aecompl}

\usepackage{graphicx}	

\title[Coorbital thermal torques]{Coorbital
  thermal torques on low-mass protoplanets}

\author[F. Masset]{Fr\'ed\'eric S. Masset\thanks{masset@icf.unam.mx}
\\
Instituto de Ciencias F\'isicas, Universidad Nacional Aut\'onoma de M\'exico, Av. Universidad s/n, 62210 Cuernavaca, Mor., Mexico}

\date{Accepted XXX. Received YYY; in original form ZZZ}

\pubyear{2017}

\begin{document}
\label{firstpage}
\pagerange{\pageref{firstpage}--\pageref{lastpage}}
\maketitle

\begin{abstract}
  Using linear perturbation theory, we investigate the torque exerted on a low-mass planet embedded in a gaseous protoplanetary disc with finite thermal diffusivity. When the planet does not release energy into the ambient disc, the main effect of thermal diffusion is the softening of the enthalpy peak near the planet, which results in the appearance of two cold and dense lobes on either side of the orbit, of size smaller than the thickness of the disc. The lobes exert torques of opposite sign on the planet, each comparable in magnitude to the one-sided Lindblad torque. When the planet is offset from corotation, the lobes are asymmetric and the planet experiences a net torque, the `cold' thermal torque, which has a magnitude that depends on the relative value of the distance to corotation to the size of the lobes $\sim\sqrt{\chi/\Omega_p}$, $\chi$ being the thermal diffusivity and $\Omega_p$ the orbital frequency. We believe that this effect corresponds to the phenomenon named `cold finger' recently reported in numerical simulations, and we argue that it constitutes the dominant mode of migration of sub-Earth-mass objects. When the planet is luminous, the heat released into the ambient disc results in an additional disturbance that takes the form of hot, low-density lobes.  They give a torque, named heating torque in previous work, that has an expression similar, but of opposite sign, to the cold thermal torque.
\end{abstract}



\section{Introduction}
The dependence on thermal diffusion of the torque experienced by a
planet embedded in a gaseous disc has been investigated mainly for
planet masses ranging from a few tens of Earth masses down to a few
Earth masses. In this mass range, a complex dependence of the torque
on thermal diffusivity has been found, which has been accounted for by
the non-linear dynamics of the corotation torque
\citep{2010ApJ...723.1393M,pbk11}. Such studies have been tackled by
means of a mixture of numerical simulations and toy models of the
coorbital region. The more direct impact of thermal diffusivity on the
linearized equations of the flow, however, has never been
investigated. Most analytic studies of the angular momentum exchange
between an external perturber and the gaseous disc have either assumed
the gas to be isothermal, or, when relaxing this barotropic
assumption, to behave adiabatically. In recent numerical simulations,
\citet{2014MNRAS.440..683L} argue for the existence of a hitherto
unmentioned component of the torque between a low-mass planet and a
gaseous disc, which they attribute to the existence of thermal
diffusion. This torque is found to originate from regions located in
the immediate vicinity of the planet, well within the length-scale of
pressure. In these regions, the gas is colder and more dense than it
would be if it behaved adiabatically. These regions are found on both
sides of corotation, and exert torques of opposite signs. Because of
an asymmetry between the torque of the inner and outer regions, they
exert a net torque. The authors dubbed this effect the `cold finger'
effect. In a different spirit, \citet{2015Natur.520...63B} have
studied the impact on migration of heat release by low-mass, luminous
planets. Thermal diffusion is naturally an essential ingredient of
such study. They find that the heat released in the vicinity of the
planet diffuses in the nearby disc and is carried away by the
Keplerian flow, yielding in steady state two hot, low-density
lobes. These lobes share a number of properties with the regions
identified by \citet{2014MNRAS.440..683L}: their characteristic size
is smaller than the length-scale of pressure, they exert antagonistic
torques on the planet, and they are asymmetric, so that they exert a
net torque on the planet. Since the regions identified by
\citet{2014MNRAS.440..683L} and those found by
\citet{2015Natur.520...63B} have similar properties but correspond to
perturbations of opposite signs (the former are dense and cold, the
latter are hot and underdense), it is not surprising that the net torques
found in these two works have opposite signs: while
\citet{2014MNRAS.440..683L} find their additional torque component to
be negative, \citet{2015Natur.520...63B} find that the release of heat
increases the total torque on the planet, up to the point that it can
become positive if the luminosity is sufficiently large. Motivated by
these findings, we undertake here the study of the torque experienced
by a low-mass planet in a disc with a finite thermal diffusivity,
using linear perturbation theory in a three-dimensional (3D) shearing sheet
(we anticipate curvature effects to be unimportant, owing to the small
size of the disturbances and their proximity to the planet). We lay
down our assumptions and write our governing equations in
section~\ref{sec:basic-equations}. We then first turn to a study of
the heat release in section~\ref{sec:effect-heat-release}. While it
may seem at first glance that dealing with the additional complexity
of heat release should be studied after the response to a cold planet,
it is actually simpler, and provides hints to the solution for a
massive, non-luminous object, which we consider in
section~\ref{sec:cold-planet}. In that section we evaluate the impact
of a finite thermal diffusivity on the torque experienced by a
low-mass, non-luminous planet. In section~\ref{sec:discussion}, we
discuss our results and compare the magnitude of the effect we found
to that of the Lindblad and corotation torques on a low-mass
planet. We draw our conclusions in section~\ref{sec:conclusion}.
\section{Basic equations}
\label{sec:basic-equations}
\subsection{Main assumptions}
\label{sec:main-assumptions}
We consider a planet of mass $M$ embedded in a protoplanetary disc on
a circular, prograde and non-inclined orbit of radius $r_p$. The
central star has a mass $M_\star$, the disc has a surface density
$\Sigma$ and an angular velocity $\Omega(r)$, where $r$ is the
distance to the central star. We assume that the disturbances that are
the subject of the present study are small compared to the pressure
length-scale~$H$ of the disc. We will assess in
section~\ref{sec:discussion} the extent to which this assumption is
justified. This assumption allows us to perform our study in the
framework of the shearing sheet \citep{shearingsheet}, here in three
dimensions\footnote{In the literature, the expression ``shearing
  box'', which would be more appropriate, almost always refer to a
  numerical device, used in particular in local magnetohydrodynamics
  simulations, rather than to the framework proposed by
  \citeauthor{shearingsheet}}. Our frame is essentially a Cartesian
box of dimensions much smaller than the planet's orbital radius, which
contains the planet and corotates with it. We use the conventional
notation for the axes: $x$ is directed along the gradient of
unperturbed velocity (i.e. along the radial direction from a global
perspective), $y$ is directed along the unperturbed motion (i.e. along
the azimuthal direction) and $z$ is perpendicular to the disc's
midplane. Although the direction of the central object is unspecified
in the shearing sheet, we will refer to the material at $x>0$ ($x<0$)
as the outer (inner) disc, implying that the central object lies on
the negative side of the $x$-axis.  The planet location is
$(x,y,z)=(x_p,0,0)$. The vanishing value of $z$ arises from the
assumption of an orbit coplanar with the disc, while $y$ can be set to
$0$ without loss of generality.

The continuity equation reads:
\begin{equation}
  \label{eq:1}
  \partial_t\rho+\nabla\cdot(\rho\mathbfit V) =0,
\end{equation}
where $\rho$ is the density and $\mathbfit V=(u,v,w)^T$ the
velocity. The Euler equation reads
\begin{equation}
  \label{eq:2}
  \partial_t\mathbfit V+\mathbfit V\cdot\nabla\mathbfit
  V+2\Omega_p\mathbfit e_z\times\mathbfit
  V=-\nabla(\Phi_t+\Phi_p)-\frac{\nabla p}{\rho},
\end{equation}
where $\mathbfit e_z$ is the unit vector along the $z$-axis, $\Omega_p$
is the rotation rate of the frame about this axis, $\Phi_p$ is the
planetary potential and $p$ is the pressure. In Eq.~\eqref{eq:2},
$\Phi_t=-q\Omega_p^2(x-x_p)^2+(1/2)\Omega_p^2z^2$ is the tidal
potential, $q$ being a dimensionless number that quantifies the shear
($q=3/2$ in Keplerian discs). Finally, the equation for the density of
internal energy $e$ reads:
\begin{equation}
  \label{eq:3}
  \partial_te+\nabla\cdot(e\mathbfit V)=-p\nabla\cdot\mathbfit
  V-\nabla\cdot\mathbfit F_H+S,
\end{equation}
where $S=S_0(\mathbfit r)+S_p(\mathbfit r)$ is a source term that
consists of the source terms $S_0$ of the unperturbed disc and
$S_p$ arising from the release of energy into the gas by
the planet, and where $\mathbfit F_H$ is the heat flux, given by
\begin{equation}
  \label{eq:4}
  \mathbfit F_H=-\chi\rho\nabla\left(\frac{e}{\rho}\right),
\end{equation}
where $\chi$ is the thermal diffusivity.  We assume the gas to be
ideal and write
\begin{equation}
  \label{eq:5}
  p=(\gamma-1)e,
\end{equation}
where $\gamma$ is the adiabatic index.  We write the perturbed
quantities as the sum of the unperturbed value and a perturbation,
denoted with a prime:
\begin{eqnarray}
  \label{eq:6}
  \rho&=&\rho_0+\rho'\\
  \label{eq:7}
  e&=&e_0+e'\\
  \label{eq:8}
  p&=&p_0+p'\\
  \label{eq:9}
  u&=&u'\\
  \label{eq:10}
  v&=&v_0+v'=-q\Omega_px+v'\\
  \label{eq:11}
  w&=&w',
\end{eqnarray}
where Eq.~\eqref{eq:10} arises from Eq.~\eqref{eq:2} with the choice
that $v_0=0$ for $x=0$ (the plane $x=0$ is therefore the planet's
corotation), which implies
\begin{equation}
  \label{eq:12}
  x_p=-\frac{\partial_xp_0}{2q\Omega_p^2\rho_0}.
\end{equation}
From now on we make the assumption that $x_p$, which is the distance
of the planet to its corotation, is much smaller than the size of the
disturbance.  We linearize Eqs.~\eqref{eq:1}-\eqref{eq:3} and
assume a steady state. We obtain
\begin{equation}
  \label{eq:13}
-q\Omega_px\partial_y\rho'+\rho_0(\partial_xu'+\partial_yv'+\partial_zw')=0
\end{equation}
\begin{equation}
  \label{eq:14}
  -q\Omega_px\partial_yu'-2\Omega
  v'=-\frac{\partial_xp'}{\rho_0}+
  \frac{(\partial_xp_0)\rho'}{\rho_0^2}-\partial_x\Phi_p
\end{equation}
\begin{equation}
  \label{eq:15}
  -q\Omega_px\partial_yv'+(2-q)\Omega_pu'=-\frac{\partial_yp'}{\rho_0}-\partial_y\Phi_p
\end{equation}
\begin{equation}
\label{eq:16}
-q\Omega_px\partial_yw'=-\frac{\partial_zp'}{\rho_0}
+\frac{(\partial_zp_0)\rho'}{\rho_0^2}-\partial_z\Phi_p
\end{equation}
\begin{eqnarray}
-q\Omega_px\partial_yp'+\gamma p_0(\partial_xu'+\partial_yv'+\partial_zw')\nonumber\\=\chi\Delta p'-\chi\frac{p_0}{\rho_0}\Delta\rho'+(\gamma-1)S_p(\mathbfit r),
\label{eq:17}
\end{eqnarray}
where we have used Eq.~\eqref{eq:5} to eliminate all instances of $e$,
and where we have assumed the size of the perturbation to be much
smaller than the length-scale over which the unperturbed quantities
vary. The second term of the right-hand side of Eq.~\eqref{eq:16} is
therefore typically smaller by a factor of order $(\lambda/H)^2$ than
the preceding term, and we neglect it. This amounts to neglecting the
vertical stratification of the disc. We take the Fourier transform of
the perturbations in $y$ and $z$, with the following conventions of
sign and normalization:
\begin{equation}
  \label{eq:18}
  \tilde\xi(x,k_y,k_z)=\iint\xi'(x,y,z)e^{-i(k_yy+k_zz)}dy\,dz
\end{equation}
\begin{equation}
\label{eq:19}
  \xi'(x,y,z)=\frac{1}{4\pi^2}\iint\tilde\xi(x,k_y,k_z)e^{i(k_yy+k_zz)}dk_y\,dk_z,
\end{equation}
where $\xi'$ represents the perturbation of an arbitrary variable, and
$\tilde\xi$ its two-dimensional Fourier transform. The system of
Eqs.~\eqref{eq:13}-\eqref{eq:17} can therefore be rewritten as the
following system of ordinary differential equations:
\begin{equation}
  \label{eq:20}
  -iqk_y\Omega_px\tilde\rho+\rho_0(\partial_x\tilde u+ik_y\tilde v+ik_z\tilde w)=0
\end{equation}
\begin{equation}
  \label{eq:21}
  -iqk_y\Omega_px\tilde u-2\Omega_p\tilde v+\frac{\partial_x\tilde p}{\rho_0}-\frac{\partial_xp_0}{\rho_0^2}\tilde\rho=-\partial_x\tilde\Phi_p
\end{equation}
\begin{equation}
  \label{eq:22}
  -iqk_y\Omega_px\tilde v+(2-q)\Omega_p\tilde u+\frac{ik_y\tilde
    p}{\rho_0}=-ik_y\tilde\Phi_p
\end{equation}
\begin{equation}
  \label{eq:23}
  -iqk_y\Omega_px\tilde w+\frac{ik_z\tilde p}{\rho_0}=-ik_z\tilde\Phi_p
\end{equation}
\begin{equation}
\begin{split}
  -iqk_y\Omega_px\tilde p&+\gamma p_0(\partial_x\tilde u+ik_y\tilde
  v+ik_z\tilde w)\\
&-\chi\Delta'\tilde p
+\frac{\chi c_s^2}{\gamma}\Delta'\tilde \rho
=(\gamma-1)\tilde S_p,
\label{eq:24}
\end{split}
\end{equation}
where we have written the forcing terms arising from the planet on the
right-hand side and where $c_s$ is the adiabatic sound speed:
\begin{equation}
  \label{eq:25}
  c_s=\sqrt\frac{\gamma p_0}{\rho_0}.
\end{equation}
In Eq.~\eqref{eq:24}, the $\Delta'$ operator is
\begin{equation}
  \label{eq:26}
  \Delta'\equiv \frac{\partial^2}{\partial x^2} - k^2,
\end{equation}
with
\begin{equation}
  \label{eq:27}
  k^2=k_y^2+k_z^2
\end{equation}
Finally, we can use Eq.~\eqref{eq:20} to eliminate the divergence of
velocity in Eq.~\eqref{eq:24} and obtain
\begin{equation}
  \label{eq:28}
  -iqk_y\Omega_px(\tilde p-c_s^2\tilde\rho)-\chi\Delta'\left(\tilde  
    p-\frac{c_s^2}{\gamma}\tilde\rho\right)=(\gamma-1)\tilde S_p.
\end{equation} 
This relationship constitutes our main equation in what follows.

In the following section, we further simplify this relation by showing
that for disturbances (i) that are not triggered by a potential and
(ii) that are much smaller than the length-scale of pressure, the
relative perturbation of pressure is negligible compared to that of
density, so that Eq.~\eqref{eq:28} can take a particularly simple
form.

\subsection{Magnitude of the perturbation of pressure}
\label{sec:magn-pert-press}
Using Eqs.~\eqref{eq:21}-\eqref{eq:23}, we can obtain the
expression of $\tilde u$, $\tilde v$ and $\tilde w$ as a function of
$\tilde p$, $\tilde \Phi_p$ and their derivatives in $x$. These can be
substituted in Eq.~\eqref{eq:20}, so as to yield an expression of
$\tilde\rho$ as a function of $\tilde p$ and $\tilde\Phi_p$. We obtain
\begin{equation}
  \label{eq:29}
  \tilde\rho-\frac{1}{D\Omega_p^2}\frac{\partial_xp_0}{\rho_0}\partial_x\tilde\rho={\cal  
    L}\left(\tilde\Phi_p+\frac{\tilde p}{\rho_0}\right) 
\end{equation} 
where the dimensionless quantity $D$ is
\begin{equation}
  \label{eq:30}
  D=q^2k_y^2x^2-2(2-q).
\end{equation}
and where the linear operator ${\cal L}$ is defined by
\begin{equation}
\begin{split}
  \label{eq:31}
  {\cal L}(Y)=&-\frac{\rho_0}{D\Omega_p^2}\partial_{x^2}^2Y
  +\frac{2\rho_0  
    q^2k_y^2x}{D^2\Omega_p^2}\partial_xY\\&
  +\rho_0\left[\frac{k_y^2}{D\Omega_p^2}\left(1-\frac{4q}{D}\right)+
    \frac{k_z^2}{q^2k_y^2\Omega_p^2x^2}\right]Y.  
\end{split}
\end{equation}  
Using Eq.~\eqref{eq:12}, Eq.~\eqref{eq:29} can be recast as
\begin{equation}
  \label{eq:32}
  \tilde\rho+\frac{qx_p}{2D}\partial_x\tilde\rho={\cal 
    L}\left(\tilde\Phi_p+\frac{\tilde p}{\rho_0}\right).
\end{equation}
Under our assumption that $x_p$ is much smaller than the typical size
of the disturbance, the second term of the left-hand side is
negligible compared to the first one since $q$ and $D$ are of order
unity, and we can write
\begin{equation}
  \label{eq:33}
  \tilde\rho \approx {\cal L}\left(\tilde\Phi_p+\frac{\tilde p}{\rho_0}\right).
\end{equation}
We now specify to the case of a perturbation not triggered by a
gravitational potential, which therefore obeys:
\begin{equation}
  \label{eq:34}
  \tilde\rho\approx {\cal L}\left(\frac{\tilde p}{\rho_0}\right).
\end{equation}
An order of magnitude of the perturbation of pressure can be obtained
by letting $k_y^{-1}\sim k_z^{-1}\sim x\sim \lambda$,
$\partial_x\tilde p\sim p/H$ and
$\partial_{x^2}^2\tilde p\sim \tilde p/H^2$, $\lambda\ll H$ being the
typical size of the density disturbance. The third and last term of
the right-hand side of Eq.~\eqref{eq:31} is then dominant and implies
\begin{equation}
  \label{eq:35}
\tilde p=O(\lambda^2\Omega_p^2\tilde\rho).
\end{equation}
As a consequence, we have
\begin{equation}
  \label{eq:36}
  \tilde p\ll H^2\Omega_p^2\tilde\rho\sim c_s^2\tilde\rho.
\end{equation}
This relation is valid for any disturbance smaller than the pressure
length-scale that verifies Eq.~\eqref{eq:34}.

\subsection{Forcing terms}
\label{sec:planet-position}
Having assumed that the distance of the planet to corotation $|x_p|$
is small compared to the size of the disturbance, we perform an
expansion to first order in $x_p$ of the planetary potential and
heating term.

The former can be obtained taking the inverse Fourier transform in $x$
of its three-dimensional Fourier transform $\bar\Phi_p(x_p,k_x,k_y,k_z)$,
which is readily obtained from Poisson's equation:
\begin{equation}
  \label{eq:37}
  \bar\Phi_p(x_p,k_x,k_y,k_z)=-\frac{4\pi GMe^{-ik_xx_p}}{k_x^2+k_y^2+k_z^2},
\end{equation}
which gives
\begin{eqnarray}
  \tilde\Phi(x_p,x,k_y,k_z)&=&\frac{1}{2\pi}\int_{-\infty}^{+\infty}\bar\Phi_p(x_p,k_x,k_y,k_z)e^{ik_xx}dk_x\nonumber\\
  \label{eq:38}
&=&-\frac{2\pi GM}{k}e^{-k|x-x_p|}.
\end{eqnarray}
In the expansion of this expression, some care must be taken that the
function $\exp (k|x-x_p|)$ does not have the same value for its left and
right derivatives at $x_p=x$. We get
\begin{equation}
  \label{eq:39}
  \tilde\Phi(x_p,x,k_y,k_z)=\Phi^{(0)}_p(x,k_y,k_z)+x_p\Phi^{(1)}_p(x,k_y,k_z)+O(x_p^2)
\end{equation}
with
\begin{equation}
  \label{eq:40}
  \Phi^{(0)}_p(x,k_y,k_z)=-\frac{2\pi GM}{k}e^{-k|x|}
\end{equation}
and
\begin{equation}
  \label{eq:41}
  \Phi^{(1)}_p(x,k_y,k_z)=-2\pi GM\mathrm{sgn}(x)
e^{-k|x|}.
\end{equation}
Similarly, specializing to a singular heat release at the planet's
location:
\begin{equation}
  \label{eq:42}
  S_p(x_p,\mathbfit r)=L\delta(x-x_p)\delta(y)\delta(z),
\end{equation}
where $\delta$ is Dirac's distribution and $L$ is the luminosity of
the planet, we obtain:
\begin{equation}
  \label{eq:43}
  \tilde S_p(x_p,x,k_y,k_z)=\tilde S_p^{(0)}(x,k_y,k_z)+x_p\tilde S_p^{(1)}(x,k_y,k_z)+O(x_p)^2
\end{equation}
with
\begin{equation}
  \label{eq:44}
  \tilde S_p^{(0)}(x,k_y,k_z)\equiv \tilde S_p^{(0)}(x)=L\delta(x)
\end{equation}
and
\begin{equation}
  \label{eq:45}
  \tilde S_p^{(1)}(x,k_y,k_z) \equiv \tilde S_p^{(1)}(x)=-L\delta'(x),
\end{equation}
where we have reduced the set of independent variables of these two
functions, as de facto they only depend on $x$.

\subsection{Decomposition of the response of the disc}
\label{sec:decomp-resp-disc}

We can formally write the linear  system of Eqs.~\eqref{eq:20}
to~\eqref{eq:24} under the concise form:
\begin{equation}
  \label{eq:46}
   S(\mathbfit Q)=\mathbfit T_\Phi+\mathbfit T_H,
\end{equation}
where
\begin{eqnarray}
  \label{eq:47}
  \mathbfit Q&=&(\rho,u,v,w,e)^T,\\
  \label{eq:48}
  \mathbfit T_\Phi&=&(0,-\partial x\tilde\Phi_P,-ik_y\tilde\Phi_p,-ik_z\tilde\Phi_p,0)^T,\\
  \label{eq:49}
  \mathbfit T_H&=&(0,0,0,0,\tilde S_p)^T 
\end{eqnarray}
are respectively the vector of the solution, and the forcing terms
arising from the planet's gravity and luminosity. In
Eq.~\eqref{eq:46}, $S$ represents the linear operator corresponding
the left-hand side  of the set of Eqs.~\eqref{eq:20}
to~\eqref{eq:24}.  Owing to the linearity of $S$, we can decompose the
solution $\mathbfit Q$ as
\begin{equation}
  \label{eq:50}
  \mathbfit Q = \mathbfit Q_\Phi+\mathbfit Q_H,
\end{equation}
where $\mathbfit Q_\Phi$ and $\mathbfit Q_H$ verify respectively
\begin{equation}
  \label{eq:51}
  S(\mathbfit Q_\Phi)=\mathbfit T_\Phi
\end{equation}
and
\begin{equation}
  \label{eq:52}
  S(\mathbfit Q_H)=\mathbfit T_H
\end{equation}
When they fulfil appropriate boundary conditions, $\mathbfit Q_\Phi$
and $\mathbfit Q_H$ characterize the disturbances excited respectively
by the planet's gravity and by the release of energy in the
surrounding gas.

We can use the expansions of section~\ref{sec:planet-position} to
further decompose the response of the disc. Using
Eqs.~\eqref{eq:43} and~\eqref{eq:49}, we can write
\begin{equation}
  \label{eq:53}
  \mathbfit T_H=\mathbfit T_H^{(0)}+x_p \mathbfit T_H^{(1)}+\mathbfit O(x_p)^2,
\end{equation}
with
\begin{eqnarray}
  \label{eq:54}
  \mathbfit T_H^{(0)}&=&[0,0,0,0,L\delta(r)]^T\\
  \label{eq:55}
  \mathbfit T_H^{(1)}&=&[0,0,0,0,-L\delta'(r)]^T.
\end{eqnarray}
The linearity of the operator $S$ implies that if we define $\mathbfit
Q_H^{(0)}$ and $\mathbfit Q_H^{(1)}$ as solutions of the linear system
respectively, with forcing terms $\mathbfit T_H^{(0)}$ and $\mathbfit
T_H^{(1)}$,
\begin{eqnarray}
  \label{eq:56}
  S[\mathbfit Q_H^{(0)}]&=&\mathbfit T_H^{(0)}\\
  \label{eq:57}
  S[\mathbfit Q_H^{(1)}]&=&\mathbfit T_H^{(1)},
\end{eqnarray}
then $\mathbfit Q_H^{(0)}+x_p\mathbfit Q_H^{(1)}$ is an expansion to
first order in $x_p$ of the solution $\mathbfit Q_H$ of
Eq.~\eqref{eq:52}. A similar decomposition can be applied to the
solution of Eq.~\eqref{eq:51}, but it will not be required in the
following.

\section{Effect of heat release}
\label{sec:effect-heat-release}
We first study the case of a luminous planet and work out the
first-order expansion $\mathbfit Q_H^{(0)}+x_p\mathbfit Q_H^{(1)}$ of
the solution $\mathbfit Q_H$ to the equation~\eqref{eq:52}. It
corresponds to a disturbance that yields a force which, by
construction, is the difference of the force exerted on a luminous
planet and the force exerted on a non-luminous planet. This
corresponds to the torque component dubbed heating torque by
\citet{2015Natur.520...63B}.  In this whole section, we do not write
an $H$ index for the different hydrodynamic variables in order to
improve legibility, but it must be understood that they are components
of $\mathbfit Q_H$.

\subsection{Advection-diffusion equation}
\label{sec:advect-diff-equat}
Since in this whole section we consider only the release of heat, the
perturbations of density and pressure verify Eq.~\eqref{eq:34}, and
Eq.~\eqref{eq:36} holds. We can therefore neglect the two occurrences
of $\tilde p$ in Eq.~\eqref{eq:28}, which takes the simple form
\begin{equation}
  \label{eq:58}
  iqk_y\Omega_px\tilde\rho+\frac{\chi}{\gamma}\Delta'\tilde\rho=\frac{\gamma-1}{c_s^2}\tilde S_p,
\end{equation}
This relation is equivalent, in real space, to
\begin{equation}
  \label{eq:59}
  \mathbfit V_0\cdot\nabla\rho'=\frac{\chi}{\gamma}\Delta\rho'-\frac{\gamma-1}{c_s^2}S_p
\end{equation}
Under the assumption $\lambda\ll H$ to which we have restricted
ourselves, the perturbation of density is therefore solution of the
simple diffusion-advection equation above, in which the advective
velocity is the unperturbed velocity of the shearing sheet, while the
pressure is essentially unperturbed ($|p'/p_0| \ll |\rho'/\rho_0|$).

We can decompose the Fourier transform of the density perturbation into
its real and imaginary parts:
\begin{equation}
  \label{eq:60}
 \tilde\rho(x,k_y,k_z)=\tilde\rho_R(x,k_y,k_z)+i \tilde\rho_I(x,k_y,k_z),
\end{equation}
where $\tilde\rho_R$ and $\tilde\rho_I$ are real
numbers. Eq.~\eqref{eq:58} is equivalent to the differential system:
\begin{equation}
  \label{eq:61}
  -q\Omega_pk_y\gamma
  x\tilde\rho_I=-\chi[\partial_{x^2}^2\tilde\rho_R-k^2\tilde\rho_R]
+\frac{\gamma(\gamma-1)\tilde S_p}{c_s^2}
\end{equation}
\begin{equation}
\label{eq:62}
  q\Omega_pk_y\gamma x\tilde\rho_R=-\chi[\partial_{x^2}^2\tilde\rho_I-k^2\tilde\rho_I].
\end{equation}
We define the dimensionless quantity $K$ as
\begin{equation}
  \label{eq:63}
  K=\frac{\chi k^3}{q\Omega_pk_y\gamma},
\end{equation}
and introduce the new variable $X$ as
\begin{equation}
  \label{eq:64}
  X=xk,
\end{equation}
which allows us to recast the system of Eqs.~\eqref{eq:61}
and~\eqref{eq:62} as
\begin{eqnarray}
  \label{eq:65}
  X\tilde\rho_I&=&K  \left(\tilde\rho_R''-\tilde\rho_R\right) 
-\frac{(\gamma-1)k\tilde S_p}{q\Omega_pk_yc _s^2}\\
\label{eq:66}
  -X\tilde\rho_R&=&K  \left(\tilde\rho_I''-\tilde\rho_I\right) ,
\end{eqnarray}
where the symbol $''$ denotes the second derivative with respect to
$X$.  The boundary conditions that the solution must satisfy are
\begin{eqnarray}
  \label{eq:67}
  \tilde\rho_R&\rightarrow& 0\mbox{~~~~when~}X\rightarrow \pm\infty \\
  \label{eq:68}
  \tilde\rho_I&\rightarrow& 0\mbox{~~~~when~}X\rightarrow \pm\infty 
\end{eqnarray}
The real part $\tilde\rho_R$ has same parity in $X$ as $\tilde S$,
whereas the imaginary part $\tilde\rho_I$ has the opposite parity.
Equations~\eqref{eq:65} and~\eqref{eq:66} describe the general response of
the gas to an arbitrary heat function $\tilde S$, when the size of the
disturbance is much smaller than the pressure length-scale.

\subsection{Force expression}
\label{sec:force-expression}
The force exerted on the planet by the perturbed density of a slab ranging from
$x_\mathrm{min}$ to $x_\mathrm{max}$ has the expression
\begin{equation}
  \label{eq:69}
  F_y=\int_{x_\mathrm{min}}^{x_\mathrm{max}}\int_{-\infty}^\infty \int_{-\infty}^\infty
\rho'\partial_y\Phi_p \,dy\,dz\,dx
\end{equation}
which can be recast, using Parseval's theorem, as
\begin{equation}
\label{eq:70}
  F_y=\frac{1}{\pi^2}\int_{x_\mathrm{min}}^{x_\mathrm{max}} \!\!\!\!dx
         \int_{k_y>0}\!\!\!\!\!dk_y\int_{k_z>0}\!\!dk_z\tilde\Phi_p
         k_y\tilde\rho_I,
\end{equation}
where $\tilde\rho_I$, as in the previous section, represents the
imaginary part of $\tilde\rho$, and where the front coefficient comes
from our conventions of Eqs.~\eqref{eq:18} and~\eqref{eq:19}. We have used
the fact that $\rho$ is even in $z$, and the fact that $\tilde\Phi_p$
is real and even in $k_y$ and $k_z$ to write the integral of
Eq.~\eqref{eq:70} over the quadrant $k_y>0, k_z>0$.

Denoting with $\tilde\rho^{(0)}$ and $\tilde\rho^{(1)}$ respectively
the density component of $\mathbfit Q_H^{(0)}$ and
$\mathbfit Q_H^{(1)}$ defined at Eqs.~\eqref{eq:56} and~\eqref{eq:57},
we can write, using the expansion of the potential of
Eq.~\eqref{eq:39}, the expansion of the force as
\begin{eqnarray}
  \label{eq:71}
  F_y=F_y^{(0)}+x_pF_y^{(1)}+O(x_p^2),
\end{eqnarray}
where
\begin{equation}
  \label{eq:72}
  F_y^{(0)}=\frac{1}{\pi^2}\int_{x_\mathrm{min}}^{x_\mathrm{max}} \!\!\!\!dx  
         \iint_{k_y>0,k_z>0}\!\!\!\!\!\!\!\!\!\!\!\!\!\!\!\!dk_y\,dk_z\tilde\Phi_p^{(0)}
         k_y\tilde\rho^{(0)}_I
\end{equation} 
and
\begin{equation}
  \label{eq:73}
  F_y^{(1)}=\frac{1}{\pi^2}\int_{x_\mathrm{min}}^{x_\mathrm{max}} \!\!\!\!\!\!dx 
          \iint_{k_y>0,k_z>0}\!\!\!\!\!\!\!\!\!\!\!\!\!\!\!\!\!\!\!\!\!dk_y\,dk_zk_y[\tilde\Phi_p^{(0)}
         \tilde\rho^{(1)}_I+\tilde\Phi_p^{(1)}
         \tilde\rho^{(0)}_I].  
\end{equation}  
When the limits of integration are
$(x_\mathrm{min},x_\mathrm{max})=(-\infty,+\infty)$, the zeroth-order
term $F_y^{(0)}$ vanishes: for symmetry reasons, there can be no net
force on a planet sitting on corotation. The evaluation of the net
force therefore requires the evaluation of the first-order term
$F_y^{(1)}$. However, it is interesting to evaluate the zeroth-order
term when the limits of integration are
$(x_\mathrm{min},x_\mathrm{max})=(0,+\infty)$, i.e. when we consider
the force exerted exclusively by the material outside corotation. Not
only does that provide one of the solutions required to evaluate the
net force ($\tilde\rho^{(0)}$), it also provides some insight into the
disc response. Using a terminology similar to the one employed for
Lindblad torques, we call this force the one-sided thermal force.

\subsection{One-sided thermal force}
\label{sec:one-sided-thermal}
As described in the previous  section, we seek here the density
response of the disc to the heating term:
\begin{equation}
  \label{eq:74}
  \tilde S^{(0)}=L\delta(x)=Lk\delta(X).
\end{equation}
We call $[R_K(X),I_K(X)]$ the solution of the differential system of
Eqs.~\eqref{eq:65} and~\eqref{eq:66} in which the forcing term is
Dirac's distribution with a unitary weight:
\begin{eqnarray}
  \label{eq:75}
  XI_K&=&K(R_K''-R_K)+\delta(X)\\
  \label{eq:76}
  -XR_K&=&K(I_K''-I_K),
\end{eqnarray}
and which satisfies $R_K(X)\rightarrow 0$ and $I_K(X)\rightarrow 0$ when
$X\rightarrow\pm \infty$. We can obtain $R_K(X)$ and $I_K(X)$ with a
shooting method, as presented in
Appendix~\ref{sec:solut-diff-syst}. We have
\begin{eqnarray}
  \label{eq:77}
  \tilde\rho_R^{(0)}(X)&=&sR_K(X)\\
  \label{eq:78}
  \tilde\rho_I^{(0)}(X)&=&sI_K(X),
\end{eqnarray}
with
\begin{equation}
  \label{eq:79}
  s=-\frac{(\gamma-1)k^2L}{q\Omega_pk_yc_s^2}
\end{equation}
Using Eqs.~\eqref{eq:40}, \eqref{eq:64}, \eqref{eq:72}, \eqref{eq:78}
and~\eqref{eq:79}, we write the force exerted by the gas at $x>0$ as
\begin{equation}
\label{eq:80}
  F_y^\text{one-sided}=\int_0^\infty\int_0^\infty f_y(k_y,k_z)dk_y\,dk_z,
\end{equation}
where the force density in Fourier space $f_y(k_y,k_z)$ is given by
\begin{equation}
  \label{eq:81}
  f_y(k_y,k_z)=\frac{2(\gamma-1)GML}{\pi q\Omega_pc_s^2}\int_0^\infty\exp(-X)I_K(X)dX.
\end{equation}
The integral of the right-hand side is a real function of the
variable $K$. We call it $F(K)$. Therefore, the one-sided force reads
\begin{equation}
  \label{eq:82}
  F_y^\text{one-sided}=\frac{2(\gamma-1)GML}{\pi q\Omega_pc_s^2}\int_0^\infty
  \int_0^\infty F(K)dk_y\,dk_z.
\end{equation}
We introduce the characteristic spatial frequency $k_c$ as
\begin{equation}
  \label{eq:83}
  k_c=\sqrt\frac{q\Omega_p\gamma}{\chi},
\end{equation}
and the dimensionless form of the wave vectors $k_y$ and $k_z$ as
\begin{eqnarray}
  \label{eq:84}
  K_y&=&k_y/k_c,\\
  \label{eq:85}
  K_z&=&k_z/k_c.
\end{eqnarray}
Eq.~\eqref{eq:63} becomes
\begin{eqnarray}
  \label{eq:86}
  K=\frac{(K_y^2+K_z^2)^{3/2}}{K_y},
\end{eqnarray}
and we rewrite Eq.~\eqref{eq:82} as
\begin{equation}
  \label{eq:87}
  F_y^\text{one-sided}=\frac{2(\gamma-1)GML}{\pi
         q\Omega_pc_s^2}k_c^2\int_0^\infty\!\!\!\int_0^\infty\!\!\!\!
         F\left[\frac{(K_y^2+K_z^2)^{3/2}}{K_y}\right]dK_ydK_z
\end{equation}
The double integral can be recast into separable form using the
variables $(\alpha,\theta)$ such that
\begin{eqnarray}
  \label{eq:88}
  K_y&=&\alpha\sin\theta\\
  \label{eq:89}
  K_z&=&\alpha\cos\theta.
\end{eqnarray}
We eventually obtain an expression involving a single integral only:
\begin{equation}
  \label{eq:90}
  F_y^\text{one-sided}=\frac{\gamma(\gamma-1)GML}{\pi 
         \chi c_s^2}\int_0^\infty F(\alpha)d\alpha. 
\end{equation}
Details about the numerical evaluation of this integral are given
in Appendix~\ref{sec:numer-eval-integr}. We find
\begin{eqnarray}
  \label{eq:91}
  \int_0^\infty F(\alpha)d\alpha\approx 0.205,
\end{eqnarray}
so that
\begin{equation}
  \label{eq:92}
  F_y^\text{one-sided}=\frac{0.0653\gamma(\gamma-1)GML}{\chi c_s^2}.
\end{equation}
This force has the same dependence on the different physical
parameters as the heating force worked out by
\citet{2017MNRAS.465.3175M} in a medium without shear, albeit with a
markedly different numerical coefficient. Remarkably, it does not
depend on the amount of shear $q$. This force is positive: the
disturbance, which corresponds to a heated region with negative
perturbation of density, tends to be displaced towards negative values
of $y$ by the Keplerian flow, and ultimately exerts a positive force
on the planet.

\subsection{Net thermal force}
\label{sec:net-thermal-force}
The thermal force exerted by the whole material, corresponding to
Eqs.~\eqref{eq:71}-\eqref{eq:73} with
$(x_\mathrm{min},x_\mathrm{max})=(-\infty,+\infty)$, requires the
evaluation of $\tilde\rho_I^{(1)}$. This quantity is given by the
solution of the differential system of Eqs.~\eqref{eq:65}
and~\eqref{eq:66} with the forcing term $\tilde S_p^{(1)}$ of
Eq.~\eqref{eq:45}
\begin{equation}
  \label{eq:93}
  S_p^{(1)}(x)=-L\delta'(x)=-Lk^2\delta'(X).
\end{equation}
We call $[r_K(X),i_K(X)]$ the solution of
Eqs.~\eqref{eq:65}-\eqref{eq:66}
in which the forcing term is the negative of the derivative of Dirac's distribution with
unitary weight:
\begin{eqnarray}
  \label{eq:94}
  Xi_K&=&K(r_K''-r_K)-\delta'(X)\\
  \label{eq:95}
  -Xr_K&=&K(i_K''-i_K),
\end{eqnarray}
which satisfies the boundary condition $r_K(X),i_K(X)\rightarrow 0$
when $X\rightarrow\pm\infty$. From Eqs.~\eqref{eq:65}, \eqref{eq:66}
and~\eqref{eq:93} we infer
\begin{equation}
  \label{eq:96}
  \tilde\rho_I^{(1)}=-\frac{(\gamma-1)k^3L}{q\Omega_pk_yc_s^2}i_K(X).
\end{equation}
Denoting with $F_y^{(1a)}$ the first part of the integral of
Eq.~\eqref{eq:73}
\begin{equation}
  \label{eq:97}
  F_y^{(1a)}=\frac{1}{\pi^2}\int_{-\infty}^{+\infty} \!\!\!\!\!\!dx 
          \iint_{k_y>0,k_z>0}\!\!\!\!\!\!\!\!\!\!\!\!\!\!\!\!\!\!\!\!\!dk_y\,dk_zk_y\tilde\Phi_p^{(0)}
         \tilde\rho^{(1)}_I,
\end{equation}
we have, using Eqs.~\eqref{eq:40}, \eqref{eq:64} and~\eqref{eq:96}:
\begin{equation}
  \label{eq:98}
  F_y^{(1a)}=\int_0^{\infty}\int_0^{\infty}f_y^{(1a)}(k_y,k_z)dk_ydk_z,
\end{equation}
with
\begin{equation}
  \label{eq:99}
  f_y^{(1a)}(k_y,k_z)=\frac{4(\gamma-1)kGML}{\pi q\Omega_pc_s^2}\int_0^{\infty}\exp(-X)i_K(X)dX.
\end{equation}
The integral of the right-hand side is a function of the variable $K$, that we
call $J(K)$. Note that we have an extra factor of $2$ in
Eq.~\eqref{eq:99} because the integration is now performed over the
whole disc, and we have used the fact that $\tilde\rho^{(1)}_I$, or
$i_K(X)$, are even functions of $X$. We have, using
Eqs.~\eqref{eq:84}-\eqref{eq:86} and~\eqref{eq:99}
\begin{multline}
  \label{eq:100}
  F_y^{(1a)}=\frac{4(\gamma-1)GML}{\pi q\Omega_pc_s^2}k_c^3\\
  \times\int_0^{\infty}\int_0^{\infty}(K_y^2+K_z^2)^{1/2}J\left[\frac{(K_y^2+K_z^2)^{3/2}}{K_y}\right]dK_ydK_z.
\end{multline}
Using again the variables of Eqs.~\eqref{eq:88}-\eqref{eq:89} and Eq.~\eqref{eq:83}, we
obtain
\begin{multline}
  \label{eq:101}
  F_y^{(1a)}=\frac{4(\gamma-1)GML}{3\pi
    q\Omega_pc_s^2}k_c^3\int_0^\infty
  J(\alpha^{2/3})d\alpha\int_0^{\pi/2}\sin^{3/2}\theta d\theta\\
\approx \frac{0.371\gamma^{3/2}(\gamma-1)GML q^{1/2}\Omega_p^{1/2}}{\chi^{3/2}c_s^2}\int_0^\infty
  J(\alpha^{2/3})d\alpha
\end{multline}
We give in Appendix~\ref{sec:numer-eval-integr} details about the
evaluation of the integral of Eq.~\eqref{eq:101}. We find
\begin{equation}
  \label{eq:102}
  \int_0^\infty
  J(\alpha^{2/3})d\alpha\approx0.616,
\end{equation}
and thus
\begin{equation}
  \label{eq:103}
  F_y^{(1a)}\approx \frac{0.228\gamma^{3/2}(\gamma-1)GML q^{1/2}\Omega_p^{1/2}}{\chi^{3/2}c_s^2}
\end{equation}
We now turn to the second part of Eq.~\eqref{eq:73}, that we call
$F_y^{(1b)}$:
\begin{equation}
  \label{eq:104}
  F_y^{(1b)}=\frac{1}{\pi^2}\int_{-\infty}^{+\infty} \!\!\!\!\!\!dx 
          \iint_{k_y>0,k_z>0}\!\!\!\!\!\!\!\!\!\!\!\!\!\!\!\!\!\!\!\!\!dk_y\,dk_zk_y\tilde\Phi_p^{(1)}
         \tilde\rho^{(0)}_I.  
\end{equation}
Using Eqs.~\eqref{eq:41}, \eqref{eq:78} and~\eqref{eq:79}, we can
write
\begin{equation}
  \label{eq:105}
  F_y^{(1b)}=\int_0^{\infty}\int_0^{\infty}f_y^{(1b)}(k_y,k_z)dk_ydk_z,
\end{equation}
with
\begin{equation}
  \label{eq:106}
  f_y^{(1b)}=\frac{4(\gamma-1)kGML}{\pi q\Omega_pc_s^2}F(K).
\end{equation}
The form $f_y^{(1b)}$ is very similar to that of $f_y^{(1a)}$, except
that it features $F(K)$ instead of $J(K)$. We can therefore directly
write
\begin{equation}
  \label{eq:107}
  F_y^{(1b)}\approx \frac{0.371\gamma^{3/2}(\gamma-1)GML q^{1/2}\Omega_p^{1/2}}{\chi^{3/2}c_s^2}\int_0^\infty
  F(\alpha^{2/3})d\alpha.
\end{equation}
An approximate numerical value of the integral (see
Appendix~\ref{sec:numer-eval-integr}) is $0.252$. Not surprisingly,
$F_y^{(1a)}$ and $F_y^{(1b)}$ have a positive sign. The former,
because the force exerted at the origin by the disturbance excited by
a planet at $x_p>0$ is positive, as the outer disc receives more heat
(we have seen that the outer disc exerts a positive force on a mass
located at the origin), and the latter because the force exerted at
the location $(x,0,0)$ (with $x>0$) by the disturbance excited by a
planet located at the origin is positive, since this location is
closer from the disturbance of the outer disc.

We eventually have
\begin{equation}
\begin{split}
  \label{eq:108}
  F_y^{(1)}=F_y^{(1a)}+F_y^{(1b)}
\approx \frac{0.322\gamma^{3/2}(\gamma-1)GML q^{1/2}\Omega_p^{1/2}}{\chi^{3/2}c_s^2}
\end{split}
\end{equation}
and the net thermal force is
\begin{equation}
  \label{eq:109}
  F_y=x_pF_y^{(1)}\approx \frac{0.322x_p\gamma^{3/2}(\gamma-1)GML q^{1/2}\Omega_p^{1/2}}{\chi^{3/2}c_s^2}.
\end{equation}
Unlike the one-sided thermal force, the net force does depend on the
shear. It also has a steeper dependence on the thermal diffusivity
than the one-sided force, and it has the same sign as $x_p$, for the
reasons explained above.

\begin{figure*}
  \includegraphics[width=.49\textwidth]{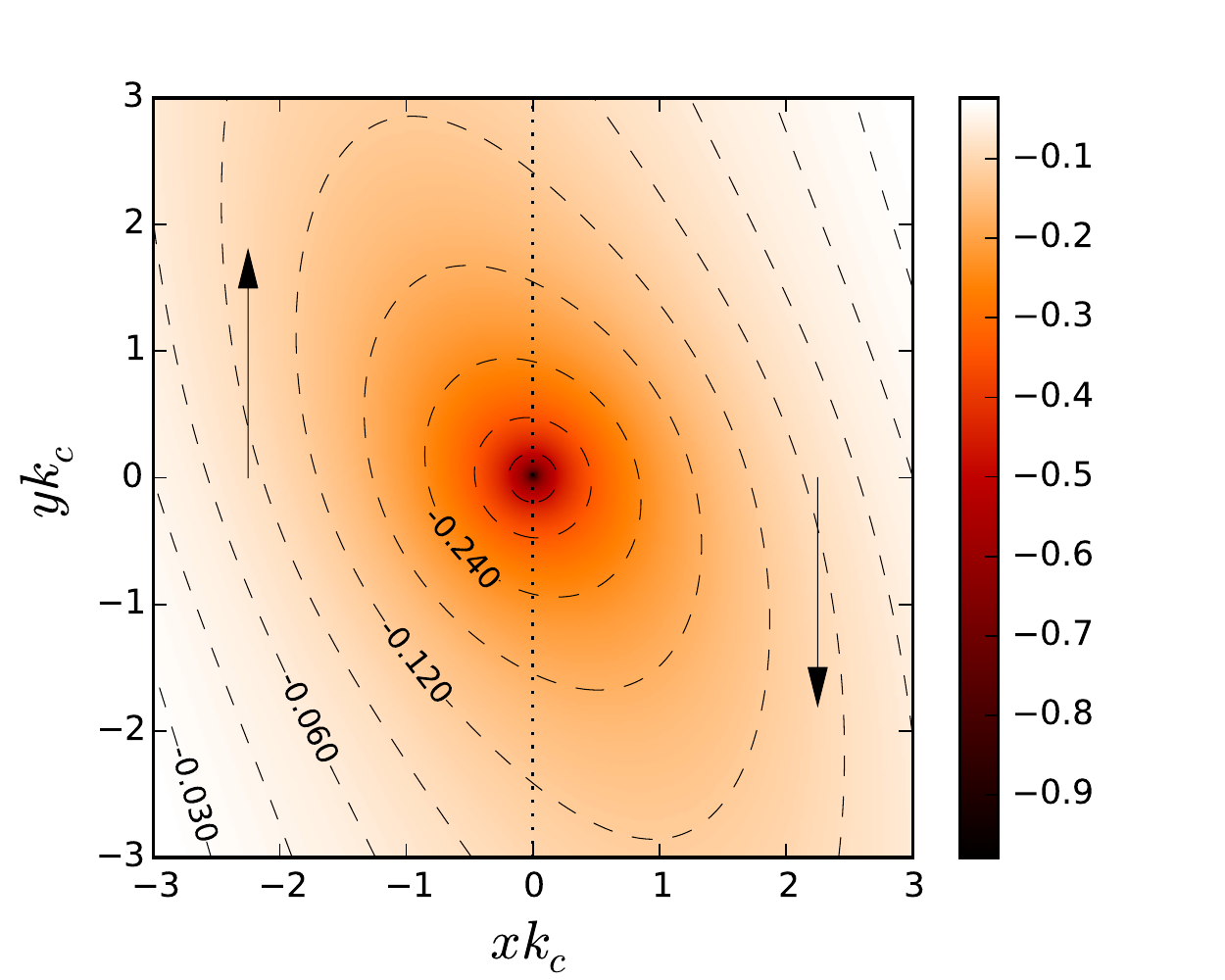}
  \includegraphics[width=.49\textwidth]{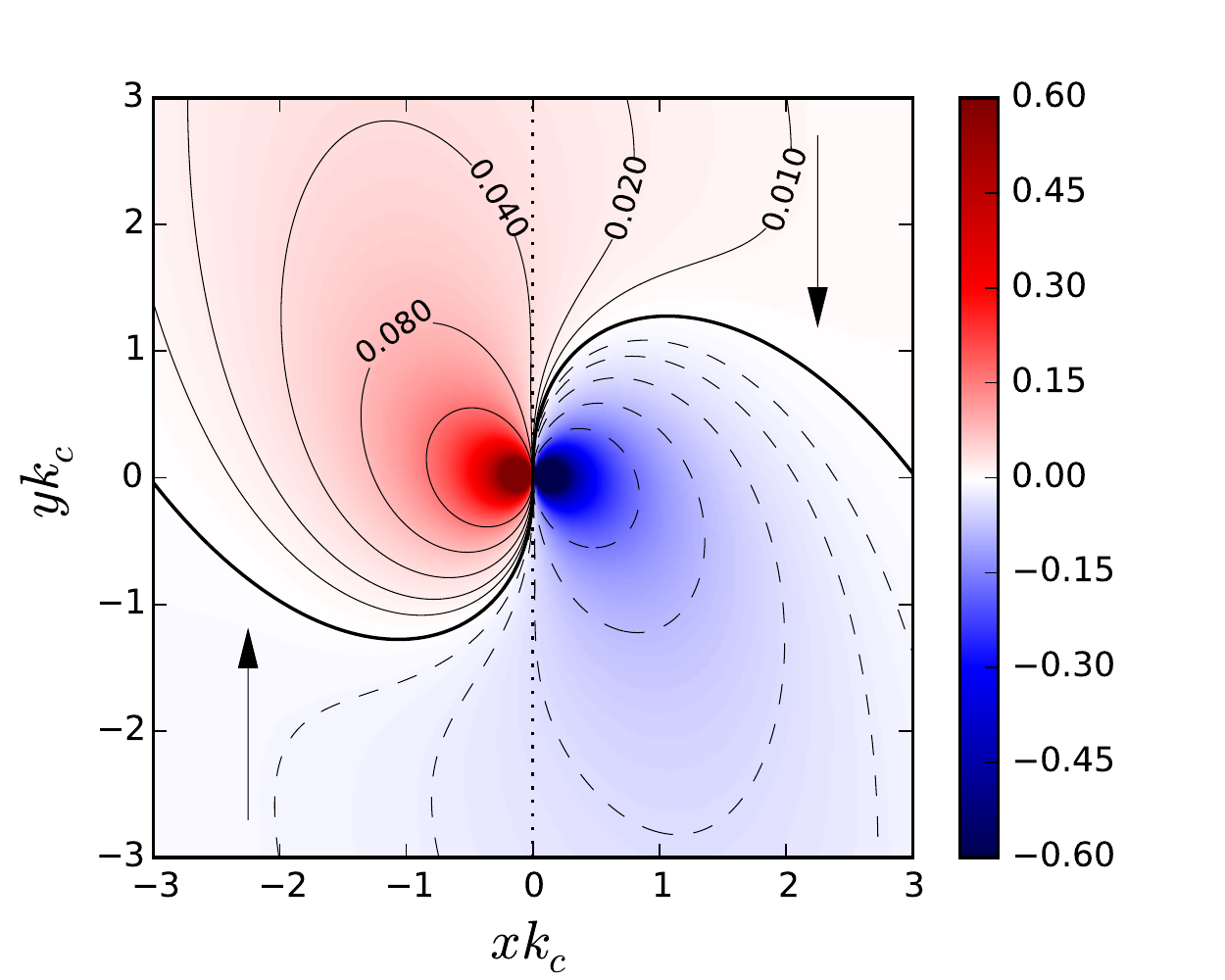}
  \caption{\label{fig:lobes01}Perturbation of surface density
    $\sigma'^{(0)}$ in units of $\gamma(\gamma-1)L/\chi c_s^2$ due to
    the singular heat release $L\delta(\mathbfit r)$ (left) and
    perturbation $\sigma'^{(1)}$ arising from the heat `dipole'
    $-L\delta'(x)\delta(y)\delta(z)$, in units of
    $\gamma(\gamma-1)Lk_c/\chi c_s^2$ (right). The map of the right
    can also be regarded as the derivative of the perturbation of
    surface density with respect to the planet position $x_p$.  These
    maps have been obtained by summing $40\,000$ Fourier components in
    geometric sequence from $k_y=10^{-4}k_c$ to $k_y=10^4k_c$.  The
    levels on the left map are in geometric sequence with a ratio of
    $\sqrt 2$ from $-3\times 10^{-2}$ to $-0.48$, while the levels on
    the right map are in geometric sequence with a ratio of $2$ from
    $\pm 1\times 10^{-2}$ to $\pm 0.16$. The thicker contour
    corresponds to the null level. The vertical arrows depict
    schematically the Keplerian flow. When the distance to the planet
    is largely smaller than $\lambda_c$ (i.e. for $|xk_c| \ll 1$ and
    $|yk_c| \ll 1$), diffusion dominates and the perturbation has
    spherical symmetry. For distances comparable to or larger than
    $\lambda_c$, advection takes over and the perturbation is
    distorted under the action of the Keplerian flow.}
\end{figure*}

\subsection{Response in real space}
\label{sec:response-real-space}
It is instructive to examine the form of the response in real
space. Denoting with a hat the one-dimensional Fourier transform in
$z$, we have
\begin{equation}
  \label{eq:110}
  \hat\rho(x,y,k_z)=\int_{-\infty}^{+\infty}\rho'(x,y,z)e^{-ik_zz}dz
\end{equation}
and
\begin{equation}
  \label{eq:111}
  \hat\rho(x,y,k_z)=\frac{1}{2\pi}\int_{-\infty}^{+\infty}\tilde\rho(x,k_y,k_z)e^{ik_yy}dk_y
\end{equation}
The perturbation of surface density $\Sigma'(x,y)$ is also
$\hat\rho(x,y,0)$. We introduce the reduced coordinates $x'=xk_c$ and
$y'=yk_c$, and the perturbation of surface density $\sigma'$ as a
function of $(x',y')$
\begin{equation}
  \label{eq:112}
  \sigma'(x',y')=\Sigma'(x'/k_c,y'/k_c).
\end{equation}
 We have
\begin{equation}
  \label{eq:113}
  \sigma'(x',y')=\frac{1}{2\pi}\int_{-\infty}^{+\infty}\tilde\rho(x'/k_c,k_y,0)e^{ik_y/k_cy'}dk_y.
\end{equation}
We can write the first-order expansion of $\sigma'(x',y')$ in $x_p$,
using the decomposition of section~\ref{sec:planet-position}. We have
\begin{equation}
  \label{eq:114}
  \sigma'(x',y')=\sigma'^{(0)}(x',y')+x_p\sigma'^{(1)}(x',y')+O(x_p^2),
\end{equation}
where $\sigma'^{(i)}$ is obtained substituting $\tilde\rho$ by
$\tilde\rho^{(i)}$ in Eq.~\eqref{eq:113}.
Noting from Eqs.~\eqref{eq:63}, \eqref{eq:79} and~\eqref{eq:83} that
for $k_z=0$, we have
\begin{equation}
  \label{eq:115}
  K=k_y|k_y|/k_c^2
\end{equation}
and
\begin{equation}
  \label{eq:116}
  \tilde\rho^{(0)}(x'/k_c,k_y,0)=s[R_K(x'k_y/k_c)+iI_K(x'k_y/k_c)]
\end{equation}
with
\begin{equation}
  \label{eq:117}
  s=-\frac{\gamma(\gamma-1)L}{\chi c_s^2}\times\frac{k_y}{k_c^2},
\end{equation}
we have
\begin{multline}
  \label{eq:118}
  \sigma'^{(0)}(x',y')=-\frac{\gamma(\gamma-1)L}{\pi\chi 
    c_s^2}\\\times \int_0^\infty K_y[R_{K_y^2}(x'K_y)\cos(y'K_y) 
  -I_{K_y^2}(x'K_y)\sin(y'K_y)]dK_y, 
\end{multline}
where we have used Eq.~\eqref{eq:84}.  In a similar fashion, using
Eq.~\eqref{eq:96}, we obtain
\begin{multline}
  \label{eq:119} 
  \sigma'^{(1)}(x',y')=-\frac{\gamma(\gamma-1)Lk_c}{\pi\chi 
    c_s^2}\\\times \int_0^\infty K_y^2[r_{K_y^2}(x'K_y)\cos(y'K_y) 
  -i_{K_y^2}(x'K_y)\sin(y'K_y)]dK_y, 
\end{multline}
where we have used the parity in $K$ of the functions $R_K$, $r_K$,
$I_K$ and $i_K$ to integrate over positive values of $K_y$ (the first
two are odd in $K$, while the other two are even in $K$).  These
relations show that the perturbation of surface density by a planet at
the origin, and its derivative with respect to the planet's distance
to corotation, are universal maps of the normalized coordinates, and
that they are proportional to the factors in front of the
integrals. The scaling factor $k_c$ is the same for the $x$- and
$y$-coordinates, which implies that the shape of the planet's response
is independent of the shear and of the thermal diffusivity. Changing
one of these parameters changes the size of the disturbance, but not
its aspect ratio. Fig.~\ref{fig:lobes01} shows the aspect of
$\sigma'^{(0)}$ and $\sigma'^{(1)}$.

\subsection{Physical picture}
\label{sec:physical-picture}
The physical picture that emerges from the previous sections is that
the heat released by a luminous planet into its surroundings yields
a low-density region, over a length-scale
\begin{equation}
\label{eq:120}
\lambda_c=k_c^{-1}=\sqrt{\chi/q\Omega_p\gamma},
\end{equation}
which is distorted by the Keplerian shear.  When the planet is centred
on corotation ($x_p=0$), no net force is exerted, for symmetry
reasons. In this situation, the outer and inner lobes exert opposite
forces of magnitude $F_y^\text{one-sided}\sim GML/\chi c_s^2$ (we
discard occurrences of the adiabatic index and numerical factors in
this discussion on orders of magnitude). When the planet is away from
corotation, the symmetry is broken and the it experiences a net force
of magnitude $F_y\sim (x_p/\lambda_c)F_y^\text{one-sided}$. Although
our expansion is valid for $|x_p|\ll \lambda_c$, we can estimate the
magnitude of the net force when the distance to corotation becomes
comparable to or larger than $\lambda_c$. \citet{2017MNRAS.465.3175M}
have evaluated the force arising from the release of heat by a
perturber in a homogeneous medium without shear. They showed that the
response time of the force, in the regime of low Mach numbers, is
$\tau\sim \chi/V^2$, where $V$ is the velocity of the perturber with
respect to the gas. In the present situation, when the response time
is shorter than the time-scale of the shear $(q\Omega_p)^{-1}$, the
shear is unimportant and the net force on the perturber can be
approximated by the expression of
\citeauthor{2017MNRAS.465.3175M}. This occurs when
\begin{equation}
  \label{eq:121}
  \frac{\chi}{(q\Omega_px_p)^2}\lesssim\frac{1}{q\Omega_p}
\end{equation}
or equivalently when $x_p\gtrsim\lambda_c$. The length-scale
$\lambda_c$ is therefore also the distance to corotation beyond which
the shear becomes irrelevant. When the planet's distance to corotation
is larger than $\lambda_c$, the force tends towards the value
$\sim GML/(\chi c_s^2)$ \citep{2017MNRAS.465.3175M}. When it is much
smaller than $\lambda_c$, it obeys the linear scaling given by
Eq.~\eqref{eq:109}. We note that Eq.~\eqref{eq:109} evaluated for
$x_p=\lambda_c$ gives approximately $2/3$ of the value given by
\citet{2017MNRAS.465.3175M}.

\section{Cold planet}
\label{sec:cold-planet}
We now turn to the case of a non-luminous planet. We denote with an
index $a$ the perturbations of the different hydrodynamics quantities
when the disc is adiabatic ($\chi=0$), and with an index $t$ the
difference between the solution with a finite thermal diffusivity and
the solution of the adiabatic case. Hence, by definition, we can
write
\begin{eqnarray}
  \label{eq:122}
  \tilde\rho&=&\tilde\rho_a+\tilde\rho_t,\\
  \label{eq:123}
  \tilde p&=&\tilde p_a+\tilde p_t.
\end{eqnarray}
In this whole section, we do not write an index $\Phi$ for the
different hydrodynamics variables in order to improve legibility, but
it must be understood that they are components of $\mathbfit Q_\Phi$,
i.e. of the disc's response to a perturber with non-vanishing
gravitational potential and with $L=0$.

Eq.~\eqref{eq:28} implies that when the disc is adiabatic ($\chi=0$)
and the planet is non-luminous ($\tilde S_p=0$):
\begin{equation}
  \label{eq:124}
  \tilde p_a-c_s^2\tilde\rho_a=0.
\end{equation}
When $\chi\ne 0$ (and $\tilde S_p=0$), Eq.~\eqref{eq:28} can be rewritten as:
\begin{equation}
  \label{eq:125}
  -iqk_y\Omega_px(\tilde p_t-c_s^2\tilde\rho_t)-\chi\Delta'\left(\tilde
    p_t-\frac{c_s^2}{\gamma}\tilde\rho_t\right)=
\chi\Delta'\left(\tilde p_a-\frac{c_s^2}{\gamma}\tilde\rho_a\right),
\end{equation}
Using Eq.~\eqref{eq:124}, we can simplify the right-hand side and obtain:
\begin{equation}
  \label{eq:126}
   -iqk_y\Omega_px(\tilde p_t-c_s^2\tilde\rho_t)-\chi\Delta'\left(\tilde
    p_t-\frac{c_s^2}{\gamma}\tilde\rho_t\right)=\frac{\gamma-1}{\gamma}\chi\Delta'\tilde p_a.
\end{equation}
The additional perturbation of density $\tilde\rho_t$ arising from a
finite thermal diffusivity obeys therefore an equation similar to
Eq.~\eqref{eq:28}, in which
$(\gamma-1)/\gamma\cdot\chi\Delta'\tilde p_a$ plays the role of the
heat source. 

Using Eq.~\eqref{eq:33}, we can write 
\begin{equation}
  \label{eq:127}
  \tilde\rho_a={\cal L}\left(\Phi_p+\frac{\tilde p_a}{\rho_0}\right)\mbox{~~~and~~~}
  \tilde\rho={\cal L}\left(\Phi_p+\frac{\tilde p}{\rho_0}\right). 
\end{equation}
We note that the thermal diffusivity $\chi$ does not feature in the
expression of the operator ${\cal L}$, so that it has same expression
in the two instances of the above identities.  The linearity of this
operator implies
\begin{equation}
  \label{eq:128}
  \tilde\rho_t={\cal L}\left(\frac{\tilde p_t}{\rho_0}\right). 
\end{equation}
Therefore, as shown in section~\ref{sec:magn-pert-press}, the relative
perturbation of pressure arising from the finite thermal diffusion is
negligible compared to that of density:
$|\tilde p_t/p_0|\ll|\tilde\rho_t/\rho_0|$, which allows us to simplify
Eq.~\eqref{eq:126}
into
\begin{equation}
  \label{eq:129}
     iqk_y\gamma\Omega_px\tilde\rho_t+\chi\Delta'
    \tilde\rho_t=\frac{\gamma-1}{c_s^2}\chi\Delta'\tilde p_a.
\end{equation}

Eq.~\eqref{eq:129} is considerably more complex than the
equation~\eqref{eq:58} for a luminous planet that we solved in
section~\ref{sec:effect-heat-release}, in which the source terms of
heat were singular at the origin. The source term in
$\Delta'\tilde p_a$ has here a complex structure over the whole wake
triggered by the planet in an adiabatic disc. It has, however, a
nearly singular component at the planet's location, as we shall see
below. We are going to restrict ourselves to the study of this
particular component, and to the force or torque it exerts on the
planet. How a finite thermal diffusivity further affects the torque
will be discussed in section~\ref{sec:addit-effect-therm}, but is not
studied in detail in this work.

We use the fact that the response of an adiabatic disc in the
immediate vicinity of a low, sub-thermal mass planet (i.e. at
distances shorter than the pressure length-scale) is such that the
planetary potential well is almost filled with enthalpy, so that the
distribution of the gas near the planet resembles that of an
isentropic atmosphere in hydrostatic equilibrium. We discuss the
validity of this assumption in
Appendix~\ref{sec:enthalpy-near-planet}. We therefore use the
approximate relationship
\begin{equation}
  \label{eq:130}
  \rho_0\Phi_p+p_a'=0
\end{equation}
to continue our calculation. Using Poisson's
equation, we arrive at
\begin{equation}
  \label{eq:131}
  iqk_y\Omega_px\tilde\rho_t+\frac{\chi}{\gamma}\Delta'\tilde\rho_t=-\frac{4\pi GM\chi\rho_0(\gamma-1)}{c_s^2\gamma}\delta(x-x_p),
\end{equation}
where we have made the approximation $\Delta(-\rho_0\Phi_p)\approx
-\rho_0\Delta\Phi_p$, that we justify in Appendix~\ref{sec:heat-source-non}.
The density perturbation arising from a finite thermal diffusivity is
formally similar to the density perturbation $\tilde\rho_H$ arising
from a singular heat release by the planet, given by
Eq.~\eqref{eq:58}, \emph{with the negative luminosity} $-L_c$, where
$L_c$ is given by
\begin{equation}
  \label{eq:132}
  L_c=\frac{4\pi GM\chi\rho_0}{\gamma}.
\end{equation}
All the results found in section~\ref{sec:effect-heat-release} can be
applied directly, except for a change of sign. Instead of two hot,
low-density lobes, we have here two cold, dense lobes on either side
of corotation. The one-sided force arising from any of these lobes
has a familiar value. Using Eqs.~\eqref{eq:92} and~\eqref{eq:132},
we find
\begin{equation}
  \label{eq:133}
\left|F_y^\text{one sided, cold}\right|=\frac{0.821(\gamma-1)G^2M^2\rho_0}{c_s^2}
\end{equation}
This quantity might be easier to recognize if we use the relationships
$\rho_0=\Sigma/\sqrt{2\pi}H$, $c_s^2=\gamma H^2\Omega_p^2$ and
write the torque $\Gamma=r_pF_y$ in terms of $\Gamma_0$ defined by
\begin{equation}
  \label{eq:134}
  \Gamma_0=\Sigma 
  r_p^4\Omega_p^2\mu^2 h^{-3},
\end{equation}
where $h=H/r_p$ is the disc's aspect ratio and $\mu=M/M_\star$.  We
obtain
\begin{equation}
  \label{eq:135}
  \left|\Gamma^\text{one sided, cold}_\text{thermal}\right|=0.33\frac{\gamma-1}{\gamma}\Gamma_0
\end{equation}
To within a dimensionless factor, this is the one-sided Lindblad
torque\footnote{The one-sided Lindblad torque depends on the disc
  profile, and is typically $\sim \Gamma_0/2$.}  \citep{w97}.  The
cold thermal lobes therefore exert on the planet torques comparable in
magnitude to the wake's torques. The analogy even holds for the signs:
the outer lobe exerts a negative torque on the planet, as does the
wake of the outer disc, while the opposite holds for the inner disc.
 
The net thermal force can be derived using Eqs.~\eqref{eq:109},
\eqref{eq:132} and~\eqref{eq:134}. We obtain:
\begin{equation}
  \label{eq:136}
  \Gamma^\text{cold}_\text{thermal}=-1.61\frac{\gamma-1}{\gamma}\frac{x_p}{\lambda_c}\Gamma_0.
\end{equation}
In Eqs.~\eqref{eq:133} and~\eqref{eq:136} we have used the subscript
\emph{thermal} to indicate that these torque components arise from
$\tilde\rho_t$, and must be added to the torque arising from
$\tilde\rho_a$ (i.e. the torque exerted on the planet in an
adiabatic disc) to get the total torque acting on the planet.

\section{Discussion}
\label{sec:discussion}
We have worked out in section~\ref{sec:effect-heat-release} the torque
arising from the perturbation $\mathbfit Q_H$, while in
section~\ref{sec:cold-planet} we have given an estimate of the torque
increment arising from a finite thermal diffusivity, with respect to
the adiabatic case. Both torques arise from the same physical process,
that of the diffusion of heat and its advection by the Keplerian flow,
over a distance typically shorter than the disc's pressure length-scale. We
give them the generic name of thermal torques. More specifically, we
keep for the former the name `heating torque' used by
\citet{2015Natur.520...63B}, and we call the latter the `cold thermal
torque' since it does not involve the release of heat by the
planet. When the planet has a finite luminosity, the total thermal
torque is the sum of the heating torque and of the cold thermal
torque.
\subsection{Comparison of the cold thermal torque and
  differential Lindblad torque}
\label{sec:comp-betw-cold}
\begin{figure*}
  \centering  
  \includegraphics[width=.7\textwidth]{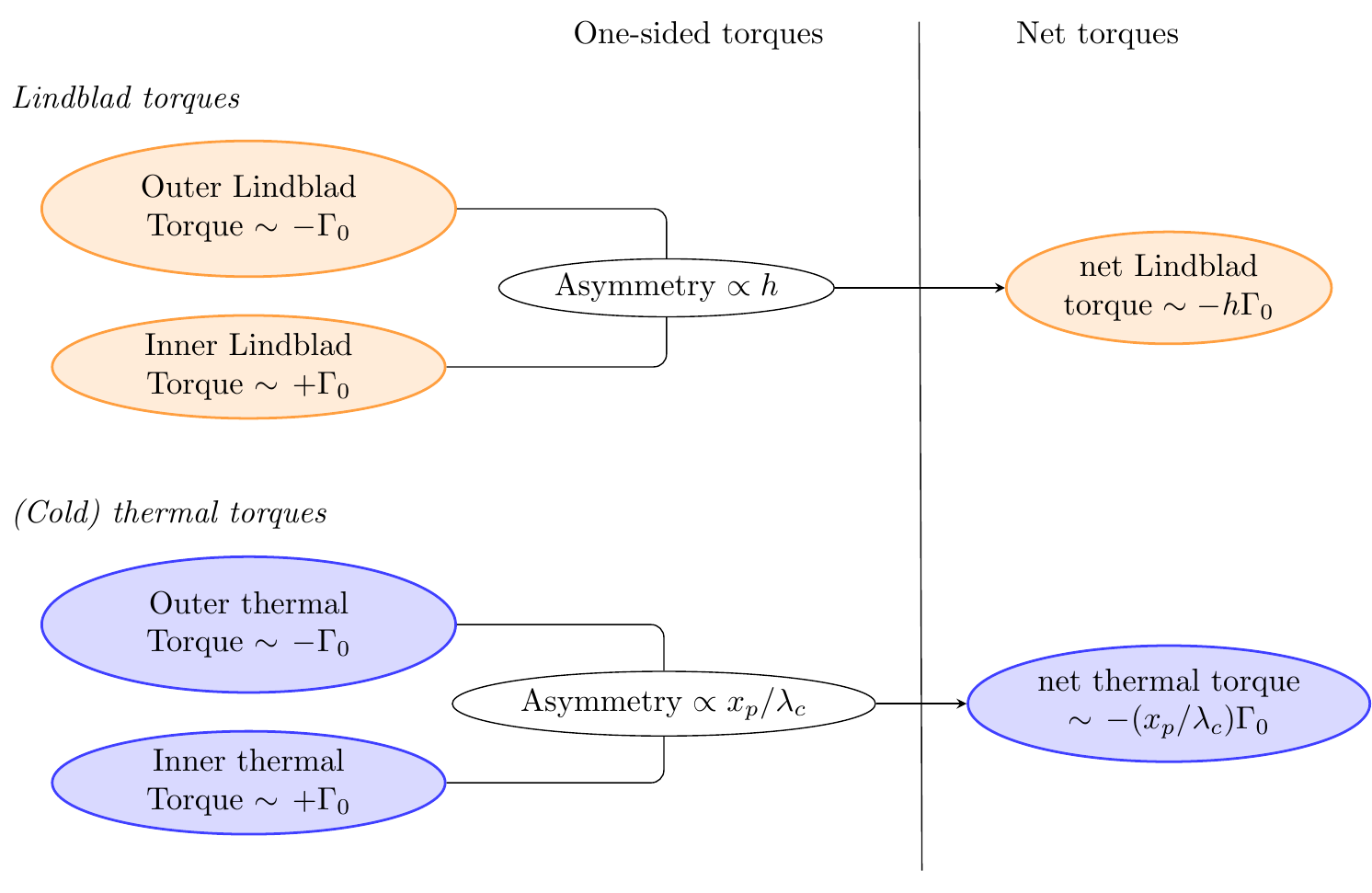}
  \caption{\label{fig:lindvsthermal}Schematic comparison between the
    Lindblad torque and the cold thermal torque. The net (or
    differential) Lindblad torque arises from an asymmetry of the
    one-sided Lindblad torques. The residual value is of order
    $-h\Gamma_0$, as it is the length-scale of pressure that sets the
    asymmetry between the torques at inner and outer Lindblad
    resonances. The corotation torque, not shown here, has same
    scaling, with a dimensionless coefficient different from that of
    the differential Lindblad torque, which depends on the disc's
    profile. Similarly, the net thermal torque arises because of an
    asymmetry between the outer and inner one-sided thermal
    torques. The amount of asymmetry, and the ultimate value of the
    net thermal torque, depends on the ratio of the corotation offset
    $x_p$ to the size of the thermal lobes $\lambda_c$. Assessing
    which of the two kinds of torque is stronger (Lindblad plus
    corotation, or thermal torque) therefore amounts to comparing
    $x_p/\lambda_c$ to $h$.}
\end{figure*} 
We have seen in the previous section that the one-sided Lindblad and
thermal torques have the same order of magnitude. Which of the net torque
(Lindblad or thermal) is larger therefore depends on the degree of
asymmetry of each torque, as depicted in
Fig.~\ref{fig:lindvsthermal}. An exploration of the parameter space
being far beyond the scope of this work, we will fix ideas using the
recent disc models of \citet{2015A&A...575A..28B}. Namely we will use
the same samples of these models as those considered by \citet[][in their
table~1]{2017MNRAS.465.3175M}. These two samples correspond to the
physical parameters in a \emph{bona fide} protoplanetary disc at
$r=3$~au (hence at distances where planetary formation is supposed to
take place) at two different dates: when the disc is young
($t=300$~kyr) and when it is more evolved ($t=1$~Myr) and has
experienced a significant drop of temperature.

The ratio of the thermal torque to the Lindblad plus corotation torque
scales with the ratio
\begin{equation}
\label{eq:137}
\frac{x_p}{\lambda_c h}=\eta\frac{H}{\lambda_c}
\end{equation}
where we have used the relation
\begin{equation}
  \label{eq:138}
  x_p=\eta h^2r_p,
\end{equation}
$\eta$ being a dimensionless coefficient of order unity that depends
on the profiles of surface density and temperature. The torque ratio
therefore scales with the ratio of the pressure length-scale (or
disc's thickness) to the size of the thermal disturbance. We have
assumed this ratio to be large in our derivation.  For the first disc
model that we consider (left column of table~1 of
\citeauthor{2017MNRAS.465.3175M}), the ratio $H/\lambda_c$ is
$\sim 4.8$, whereas for the second disc model it has the value
$\sim 12$. This essentially validates the key assumption made in
section \ref{sec:main-assumptions} that $\lambda_c\ll H$.

We can make more quantitative the comparison between the Lindblad plus
corotation torque and the cold thermal torque by taking into account
the dimensionless factors of order unity.  When the surface density
and temperature profiles are power laws of the radius (respectively
with exponents $-\alpha$ and $-\beta$), one has, using Eq.~\eqref{eq:12}
\begin{equation}
  \label{eq:139}
  \eta=\frac{\alpha}{3}+\frac{\beta+3}{6}. 
\end{equation}
In the disc models that we consider, in which
$(\alpha,\beta)\approx(1/2,1)$ for the first model and
$(\alpha,\beta)\approx (0,1)$ for the second model, we have
respectively $\eta\approx 0.8$ and $\eta\approx 0.7$.
Specialising, from now on, to the case $\gamma=7/5$, adequate for
protoplanetary discs, we get from Eq.~\eqref{eq:136}:
\begin{eqnarray}
  \nonumber
  \Gamma^\text{cold}_\text{thermal}&\approx& -1.8h\Gamma_0\mbox{~~~for
                                             model~1,}\\
  \nonumber
  \Gamma^\text{cold}_\text{thermal}&\approx& -3.9h\Gamma_0\mbox{~~~for model~2.}
\end{eqnarray}
These estimates are to be compared with the typical value of the
Lindblad plus corotation torque\footnote{We warn the reader against a
  possible confusion between our definition of $\Gamma_0$, which
  scales as $h^{-3}$ as does the one-sided Lindblad torque, and the
  definition adopted in many publications, which involves $h^{-2}$
  rather than $h^{-3}$, and which corresponds to the scaling of the
  differential Lindblad or corotation torques. This is why in the
  present work the net Lindblad or corotation torque scales with
  $h\Gamma_0$ rather than $\Gamma_0$.}, which is generally in the
range $[-2h\Gamma_0/\gamma,2h\Gamma_0/\gamma]$ \citep[see
e.g.][]{2015MNRAS.452.1717L,2017arXiv170708988J}, and rather on the
lower side of this interval for planets of the order of one to a few
Earth masses.

This indicates that for protoplanets or planetary embryos with a mass
sufficiently small to be subjected to the thermal torque, the latter
is an essential component of the total torque. In particular, in discs
with low thermal diffusivity (as the disc model~2), the thermal torque
can be so large that it makes the Lindblad and corotation torques
virtually irrelevant. Since the thermal torque scales with the offset
to corotation $x_p$, embryos mainly driven by the thermal torque
should be trapped at or near the pressure traps (locations where the
pressure gradient, and therefore the offset to corotation, cancel out)
much as lower mass objects driven by aerodynamic drag.

\subsection{Critical mass for heat release}
\label{sec:mass-range-validity}
The planetary mass up to which thermal torques are significant is an
open question. In the context of dynamical friction,
\citet{2017MNRAS.465.3175M} argue that the estimate of the heating
force obtained by a linear analysis holds when the heat released by
the planet entirely ends up as an excess of internal energy outside of
the Bondi sphere\footnote{If the luminosity is not too large, the
  Bondi sphere encloses the region where the flow is
  non-linear.}. This occurs when the time-scale for heat diffusion
across Bondi's radius $r_B=GM/c_s^2$ is shorter than the acoustic time
$r_B/c_s$. This condition translates into $M<M_c=\chi c_s/G$. 
For masses larger than this critical value, it is not guaranteed that
the internal energy injected in the gas near the planet
emerges as an excess of internal energy outside of the Bondi sphere,
and one may expect a cut-off of the thermal force.

The exploration of the parameter space of \citet{2015Natur.520...63B},
which shows a cut-off of the heating torque above a few Earth masses,
is compatible with this expectation. Similarly, the work of
\citet{2014MNRAS.440..683L}, which is likely the only numerical
evidence of the cold thermal torque to date (as we shall see in
section~\ref{sec:cold-thermal-torque}), shows that this effect
vanishes past $\sim 3M_\oplus$. This set of evidence is rather slim,
however, and a thorough study of the thermal torques as a function of
the planetary mass is necessary to provide useful formulae that can be
incorporated to models of planetary population synthesis. Owing to the
non-linear nature of the flow within the Bondi sphere, a systematic
study of the mass dependence may need to resort to numerical
simulations. There is nevertheless a foreseeable difficulty inherent
to such study. The mass range of interest and the resolution
requirements are such that the time step of an explicit scheme will
not be limited by the sound speed, but necessarily by heat diffusion,
potentially yielding very short time steps. Unless heat diffusion is
dealt with in an implicit manner, or accelerated by the use of a super
time-stepping technique \citep{CNM:CNM950}, the calculations may prove
impracticable.

We also note that even in the barotropic case complex phenomena occur
in the Bondi sphere of a low-mass planet
\citep{2015arXiv150503152F,2015MNRAS.446.1026O,2015MNRAS.447.3512O,2017AJ....153..124F}. In
the particular case in which the gas is isothermal, there is a large mass
build-up within the Bondi sphere owing to the lack of compressional
heating, with potentially a large impact on the torque
\citep{2015arXiv150503152F}. For sub-critical planets ($M<M_c$), the
flow may be considered as nearly isothermal within the Bondi sphere. The
enhancement of density $\rho_t$ experienced in such case by a cold
planet leads to a similar mass build up as in the isothermal case, which
may have an impact on the torque. The investigation of this highly
non-linear small-scale flow is largely beyond the scope of this
work. It likely requires to be tackled by means of numerical
simulations. While the Bondi sphere of nearly thermal-mass planets
($\mu\sim h^3$) can be resolved on modern computational platform, that
of deeply embedded objects ($\mu \ll h^3$) such as those considered
here cannot be resolved with present-day computational resources
\citep{2015arXiv150503152F}, at least for global disc simulations.

Until a detailed exploration of the thermal effects that resolves the
flow at the sub-Bondi scale and provides the magnitude of the force as
a function of the planetary mass is available, we caution that
estimates of the thermal torques should be valid only when
$M<M_c$, and regarded as upper values otherwise.

\subsection{Comparison to earlier work}
\label{sec:comp-earl-work}
\subsubsection{Cold thermal torque and the `cold finger' effect}
\label{sec:cold-thermal-torque}
The cold thermal torque has been quite elusive so far in numerical
simulations of embedded protoplanets. There are several reasons for
that: (i) it requires a finite thermal diffusivity of the gas, while
the vast majority of the numerical studies used either isothermal or
adiabatic setups; (ii) a finite thermal diffusivity is generally
achieved through the use of some sort of radiative transfer (such as
flux limited diffusion), which considerably increases the numerical
cost over what could be achieved if thermal diffusion was modelled as
in the present work, and consequently decreases the size of the
parameter space that can be explored, (iii) thermal torques are
exerted on low or very low mass planets (up to a few Earth masses at
most), which are not systematically included in studies of planet-disc
interactions with radiative transfer and (iv) the thermal lobes are
typically an order of magnitude smaller than the length-scale of
pressure, and are barely captured even with state-of-the-art
resolutions. To the knowledge of the author, there is only one mention
in the literature of the effect that we described here as the cold
thermal torque: \citet{2014MNRAS.440..683L} found that Earth-sized
planets in 3D radiative discs were subjected to a negative torque
which could be accounted for neither by the Lindblad torque nor by the
corotation torque. The radial density of the torque shows strong
contributions bound to the coorbital region, negative outside and
positive inside, with a marked asymmetry in favour of the outside
component. All of these features are compatible with the effect we
report here. One can furthermore estimate the thermal diffusivity at
the midplane of the disc considered by these authors to be
$\chi\approx 1.5\times 10^{15}$~cm$^2$s$^{-1}$, from which we can
estimate, for their planets orbiting a solar mass star at
$r_p=5.2$~au, that $\lambda_c=0.014$~au, compatible with the extent of
the peaks found in their radial torque distribution. Incidentally,
this shows that in their disc, they have $H/\lambda_c\approx 14$,
bringing further evidence that the thermal lobes are in general much
smaller than the pressure length-scale. We can also estimate the
corotation offset $x_p$, in their setup, to be of the order of
$6\times 10^{-3}$~au, which represents a fair fraction of $\lambda_c$,
the characteristic size of the thermal lobes, and can account for the
strong asymmetry between the inner and outer torques. Finally, they
observe a strong difference in the torque distribution between their
lightest planet (with mass $2M_\oplus$) and the next one (with mass
$3M_\oplus$), indicating that the effect is probably already cut-off
in their simulations, at least above the lowest mass considered. As
further evidence to support this claim, we note that
Eq.~\eqref{eq:136} predicts a cold thermal torque of the order of
$-5.5h\Gamma_0$, which would result in a total torque much larger in
absolute value than the values they report even for $2M_\oplus$,
which seems to indicate that even for this lowest mass the effect is
already significantly cut-off. We comment however that this
discrepancy can also be attributed, at least partially, to the low
resolution with which the lobes are captured.

These authors reported their effect as the formation of `cold
fingers'. This denomination likely arises from subsequent
two-dimensional simulations that they perform, in which the cooling of
fluid parcels is not due to heat diffusion but to an exponential
relaxation towards a prescribed equilibrium temperature, resulting in a
larger entropy loss for gas parcels that experience more compressional
heating. This numerical experiment is enlightening as it explains the
formation of cold, dense structures past the planet. The resulting
disturbances, nonetheless, do not arise from an advection-diffusion
equation as the one we solved here, but merely from advection and
cooling, which results in much more narrow, elongated features. We
suggest that the denomination `cold lobes' is more adequate when
there is thermal diffusion.

\subsubsection{Heating torque} \label{sec:heating-torque} The heating
torque, analysed in section~\ref{sec:effect-heat-release}, has been
studied by means of numerical simulations by
\citet{2015Natur.520...63B}. Most of the remarks that we made for the
work of \citet{2014MNRAS.440..683L} in
section~\ref{sec:cold-thermal-torque} also apply to this work, except
for a change of sign. Also, the disc's thermal diffusivity in this
work is approximately $2.9$~times larger than that of
\citeauthor{2014MNRAS.440..683L}, which implies that the length-scale
$\lambda_c$ of the thermal lobes is $1.7$~times larger (the distance
to the central star, and the mass of the latter, are the same in both
works). This is compatible with the radial torque density presented by
\citet{2015Natur.520...63B}. We mention that their `neutral run', in
which the heat release is disabled, does include a cold thermal
torque. The planet's luminosity in the fiducial calculation of this
work is $L=6.0\times 10^{27}$~erg~s$^{-1}$. Noting that
Eq.~\eqref{eq:109} can be rewritten, using Eq.~\eqref{eq:138}
as
 \begin{eqnarray}
  \label{eq:140}
  \Gamma^\text{heating}=0.261\eta\frac{GML}{(\chi\Omega_p)^{3/2}},
\end{eqnarray}
and noting that here $\eta=5/6$, we would predict for the heating
torque a value of
$\Gamma^\text{heating}\sim 2.7\times 10^{36}$~g~cm$^2$s$^{-2}$. This
value is typically one order of magnitude larger than the value
measured. As for the simulations of \citet{2014MNRAS.440..683L}, this
may constitute an indication that the thermal torques are cut-off. The
discrepancy may also partly arise from the barely sufficient
resolution ($\lambda_c$ is just twice the radial resolution, and
marginally smaller than the azimuthal resolution in that work).

The main trend found by \citet{2015Natur.520...63B}, which is a strong
dependence on the disc's opacity, is qualitatively compatible with the
results of Eq.~\eqref{eq:109}: the larger the opacity, the smaller the
thermal diffusivity, and the larger the heating torque. The comparison
cannot be made quantitative, however: the fiducial mass considered in
this work is largely beyond the critical mass $\chi c_s/G$, and the
latter furthermore varies when the opacity varies. They also find the
heating torque to scale with the distance to corotation, which is
compatible with our findings.

Finally, we mention that the luminosity $L_c$ of Eq.~\eqref{eq:132} is
$\sim 1.1\times 10^{27}$~ergs~s$^{-1}$, largely smaller than the
luminosity of the fiducial run of \citet{2015Natur.520...63B}. It
would correspond to a mass doubling time of $\sim500$~kyr. The cold
thermal torque has therefore only a mild impact in the runs of that
work.

\subsection{Additional effect of thermal diffusion on the torque}
\label{sec:addit-effect-therm}
As mentioned in section~\ref{sec:cold-planet}, we have restricted
ourselves, in the cold case, to the study of the disturbance triggered
by the singular component at the planet's location. Yet, one may
expect thermal diffusion to have a more direct, intuitive effect on
the wake's torque, if we omit the singular component: at low thermal
diffusivity, the wake essentially behaves adiabatically and exerts
the same torque as in an adiabatic disc, whereas at large thermal
diffusivity it should behave isothermally and exert the same torque
as in an isothermal disc. It is easy, however, to realize that the
thermal diffusivity is usually sufficiently small for the wake's
torque to have the adiabatic value, when, as we noted in previous
sections, $H \gg \lambda_c$. The exchange of angular momentum between
the planet and the disc at a given Lindblad resonance occurs
essentially over the first wavelength of the wave launched at the
resonance. For waves with low azimuthal wavenumber $m$
($m < h^{-1}$), the wave vector of the disc's response over the first
wavelength is dominated by its radial component and has the order of
magnitude $|\mathbfit k|\sim (m/h^2)^{1/3}/r_p$, whereas waves with
large azimuthal wavenumber ($m>h$) have a wave vector dominated by
its azimuthal component so that $|\mathbfit k|\sim m/r$. A wave with
wave vector $k$ and frequency $\omega$ behaves adiabatically if $k$ is
smaller than
\begin{equation}
  \label{eq:141}
  k_\text{cut}=\sqrt{\omega/\chi}
\end{equation}
The frequency of the waves launched at Lindblad resonances is the
local epicyclic frequency, which is also close to $\Omega_p$ for a
Keplerian disc. Waves with low azimuthal wavenumber therefore satisfy
$k<(\lambda_c/H)k_\text{cut}$, and behave adiabatically. Waves with
high azimuthal wavenumber will reach the cut-off wavenumber for
$m\sim r/\lambda_c=(H/\lambda_c) h^{-1}$. This wavenumber is
considerably larger than the wavenumber for the torque peak, which
occurs for $m\sim h^{-1}/2$ \citep{w97}, so that virtually all
resonances involved in the angular momentum exchange should excite an
adiabatic response. From these arguments we also see that a transition
towards another regime should be expected when $H\sim \lambda_c$,
i.e. when $\chi \sim H^2\Omega_p$. This is the critical thermal
diffusivity considered by \citet{2010ApJ...723.1393M}, who performed a
fit of the Lindblad torque as a function of the thermal
diffusivity. Alternatively, \citet{pbk11} use a one-dimensional wave
model in a uniform medium as a guideline, and work out an effective
adiabatic index $\gamma_\mathrm{eff}$. They find a turnover
diffusivity a factor of $\sim h$ smaller than $H^2\Omega_p$, as they
consider the wave frequency in the inertial frame ($m\Omega_p$) rather
than in the local frame. It is unclear what value one should expect
for the Lindblad torque when $\chi$ becomes a sizeable fraction of
$H^2\Omega_p$. In this regime, the waves launched at resonances are
damped near their region of excitation. This process has been studied
by \citet{1996ApJ...472..789C} when thermal diffusion arises from
radiative transfer. Although these issues definitely require further
work, it should be clear that, for values of the thermal diffusivity
typical of protoplanetary discs, they correspond to minute corrections
to the torque, in comparison to the large effect of the cold thermal
torque.

\subsection{A simple expression for the total thermal torque}
\label{sec:total-thermal-torque}
The total thermal torque is the sum of the heating torque and of the
cold thermal torque:
\begin{equation}
  \label{eq:142}
  \Gamma^\mathrm{total}_\mathrm{thermal}=\Gamma_\mathrm{thermal}^\mathrm{heating}+\Gamma_\mathrm{thermal}^\mathrm{cold},
\end{equation}
while the total torque acting on the planet is
\begin{equation}
  \label{eq:143}
  \Gamma^\mathrm{total}=\Gamma^\mathrm{total}_\mathrm{thermal}+\Gamma_\mathrm{adiabatic}
\end{equation}
The heating torque given by Eq.~\eqref{eq:109} can be cast in a simple
form using the critical luminosity $L_c$ of Eq.~\eqref{eq:132}. We obtain
\begin{equation}
  \label{eq:144}
  \Gamma^\mathrm{heating}_\mathrm{thermal}=1.61\frac{\gamma-1}{\gamma}\frac{x_p}{\lambda_c}\frac{L}{L_c}\Gamma_0,
\end{equation}
while the total thermal torque is
\begin{equation}
  \label{eq:145}
  \Gamma^\mathrm{total}_\mathrm{thermal}=1.61\frac{\gamma-1}{\gamma}\frac{x_p}{\lambda_c}\left(\frac{L}{L_c}-1\right)\Gamma_0.
\end{equation}
This expression can also be recast under the convenient form
\begin{equation}
  \label{eq:146}
  \Gamma^\mathrm{total}_\mathrm{thermal}=1.61\frac{\gamma-1}{\gamma}\eta\left(\frac{H}{\lambda_c}\right)\left(\frac{L}{L_c}-1\right)
  h\Gamma_0,
\end{equation}
where $\eta$ is given by Eq.~\eqref{eq:139}, $\lambda_c$ by
Eq.~\eqref{eq:120}, $L_c$ by Eq.~\eqref{eq:132} and $\Gamma_0$ by
Eq.~\eqref{eq:134}.

\subsection{Dependence on the disc gradients}
\label{sec:depend-disc-grad}
The analysis presented in
Sections~\ref{sec:basic-equations}-\ref{sec:cold-planet} allows for a
radial gradient of density, but assumes the background temperature to
be uniform. The perturbation of density in the heated region does not
depend on the background density [a similar result was obtained by
\citet{2017MNRAS.465.3175M} for the hot plume created by a hot body in
a uniform medium]. This result holds as long as the perturbation of
density is a small fraction of the unperturbed density. In the cold
case, the perturbation of density does feature the unperturbed density
at the planet's location. Therefore, no dependence of the thermal
torques on the unperturbed density gradient should be expected, other
than through the dependence on the corotation offset $x_p$.

We can estimate the dependence of the thermal torque on the
temperature gradient as follows. A temperature gradient yields a
correction that has the order of magnitude:
\begin{equation}
  \label{eq:169}
  \Delta_\beta\Gamma_\mathrm{thermal}\sim
  \Gamma^\mathrm{one-sided}c_s^2\partial_r\left(\frac{1}{c_s^2}\right)\lambda_c=\beta\Gamma^\mathrm{one-sided}\frac{\lambda_c}{r_p}.
\end{equation}
We compare this correction to the thermal torque
$\Gamma_\mathrm{thermal}\sim
(x_p/\lambda_c)\Gamma^\mathrm{one-sided}$:
\begin{equation}
  \label{eq:171}
  \left|\frac{\Delta_\beta\Gamma_\mathrm{thermal}}{\Gamma_\mathrm{thermal}}\right|\sim|\beta|\frac{\lambda_c^2}{r_px_p}\sim|\beta|\frac{\lambda_c^2}{H^2}\ll
  1.
\end{equation}
The correction is therefore negligible. Contrary to what happens for
the case of the Lindblad torque, the region that exerts the torque is
so compact that the gradient of background temperature does not have a
sizeable impact on the torque value, and the thermal torques have no
dependence on the temperature profile other than the one borne by the
corotation offset $x_p$ (see equations~\ref{eq:138} and~\ref{eq:139}).

The torque expressions given in section~\ref{sec:total-thermal-torque}
are therefore valid in discs with arbitrary surface density and
temperature gradients.

\subsection{On the negative luminosity of the cold case}
\label{sec:universality-l_c}
\begin{figure*}
  \centering 
  \includegraphics[width=\textwidth]{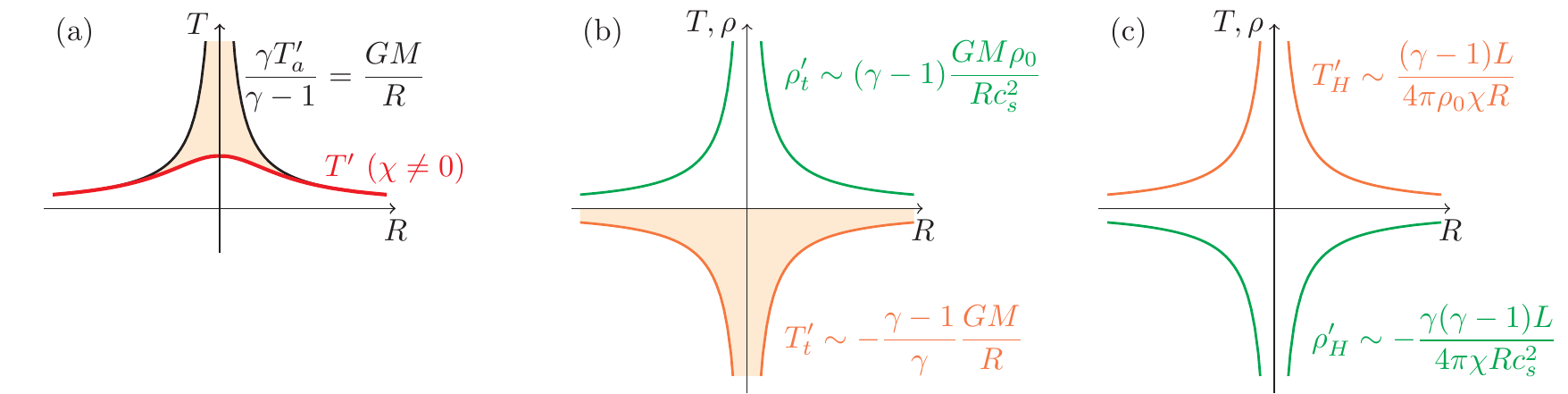}
  \caption{\label{fig:neglum}Schematic representation of the
    perturbation of temperature or density in the vicinity of the
    planet in various cases. The variable $R$ denotes here the
    distance to the planet in an arbitrary direction. Panel~(a) shows
    the perturbation of temperature $T_a'$ in an adiabatic disc (black
    curve). The form of this perturbation comes from the fact that the
    perturbation of enthalpy $\gamma T_a'/(\gamma-1)$ is nearly the
    opposite of the planetary potential. When thermal diffusion is
    introduced, the temperature profile is flattened (red curve). The
    new perturbation of temperature $T'$ can then formally be written
    as $T_a'+T_t'$, where $T_t'<0$ corresponds to the coloured area,
    and tends to the negative of $T_a'$ when $R\rightarrow
    0$. Panel~(b) shows the graph of $T_t'$, and of the associated
    perturbation of density. Since the perturbation of pressure is
    small, we have $\rho'/\rho_0+T'/T_0\approx 0$, or
    $\rho'\approx -\gamma\rho_0T'/c_s^2$. The shape of $\rho'_t$ is
    therefore the negative of the potential, and it scales here with
    $R^{-1}$ because the planetary potential does. Panel~(c) shows the
    disturbances of density and temperature imparted by a luminous,
    massless perturber of luminosity $L$, which correspond the heating
    torque. Here $\rho'_H$ scales with $R^{-1}$ because it obeys a
    diffusion equation in three dimensions around a point-like
    source. The comparison of the scaling of $\rho'_H$ and $\rho'_t$
    (green curves in the electronic version) shows that the latter
    corresponds to the perturbation induced by a singular massless
    heat sink with the luminosity $-L_c$, where $L_c$ is given by
    Eq.~\eqref{eq:132}. }
\end{figure*}

The physical picture of the heating torque gives some insight into the
mechanism of the cold thermal torque, which is very similar in
nature. When the disc is adiabatic, there is a peak of temperature
$T=p/\rho$ at the location of a non-luminous planet, since the
potential well of the planet is almost topped off with enthalpy. The
introduction of thermal diffusion flattens this peak over a distance
$\sim\lambda_c$. As for the heating torque, this is achieved with
perturbations of temperature and density of opposite signs, and
virtually no perturbation of pressure. This situation is represented
graphically in Fig.~\ref{fig:neglum}, which shows the perturbations of
temperature and density at distances from the planet sufficiently
small so that they can be regarded as having spherical symmetry.
We see in this figure that the density perturbation associated with the
finite thermal diffusivity seems to stem from a singular heat source
with negative luminosity. At larger scale, in steady state, the
perturbed density adopts the same pattern as the perturbed density
associated with the heat release, with an opposite sign. This shows that
the cold thermal torque and the heating torque are two slightly
different versions of the same process of diffusion-advection.

The fact that the density perturbation associated with thermal diffusion
seems to stem from a point-like source arises from a coincidence
between two laws in $R^{-1}$: that of the planetary potential and
that of diffusion in three dimensions around a point-like
source\footnote{A simple manner to obtain the law of $\rho'_H$ in the
  immediate vicinity of the planet (i.e. at distances much shorter
  than $\lambda_c$) consists in solving Eq.~\eqref{eq:59} in which we
  neglect the left-hand side, which represents the advective
  term.}. This coincidence is not fortuitous: it simply arises from
the fact that both processes are described by two formally similar
Poisson's equations with a singular right-hand side. This coincidence
does not hold in two dimensions, where the planetary potential is
still represented by a $R^{-1}$ law, whereas the perturbation of
density arising from the heat release by a point-like source is
solution of a two-dimensional Poisson's equation, with a divergence in
$\log R$ in the vicinity of the perturber. In these circumstances, the
perturbation of density arising from thermal diffusion near a cold
planet cannot be regarded as originating from a point-like
source. More generally, this underscores the need for
three-dimensional calculations to reproduce correctly the properties
of the thermal torques.

This mechanism (flattening of the temperature peak by thermal
diffusion, which triggers a density perturbation similar to the one
that would arise from a singular heat sink) should be valid every time
a point-like mass moves at low Mach number within a non-adiabatic
gas. Under these circumstances, thermal diffusion imparts a
perturbation to the flow in addition to the adiabatic response, which
is the same as the perturbation that would be imparted by a
\emph{massless heat sink with the negative luminosity}
$-4\pi GM\chi\rho_0/\gamma$. Naturally, despite what this formulation
might suggest, there is no net heat flux on to the perturber: the
temperature gradient vanishes in the vicinity of the latter, and so
does the heat flux.

Recently, \citet{2017arXiv170401931E} have studied the evolution of
the eccentricity and inclination of low-mass planets in radiative
discs. They found that luminous planets experience a growth of these
orbital elements. They interpret these findings as arising from the
force exerted by a hot, low-density region trailing the planet on its
epicyclic or vertical motion. They also found that non-luminous
planets, when embedded in radiative discs, experience a faster decay
of eccentricity and inclination than in an adiabatic disc. These
findings are compatible with these planets having on the contrary a
cold, dense trail that contribute to damp these orbital elements, in
addition to the well-known action of the coorbital Lindblad resonances
\citep{arty93b} and coorbital vertical resonances
\citep{1994ApJ...423..581A}, much as we would expect if the planet
imparted on the flow an additional perturbation similar to that
arising from a heat sink. We also note that
\citet{2017arXiv170401931E} used the same disc parameters as
\citet{2015Natur.520...63B}. Their fiducial planet has therefore the
same, small value of $L_c$ (see section \ref{sec:heating-torque}), and
all the planetary luminosities that they consider exceed $L_c$,
explaining why all their luminous planets have finite eccentricity and
inclinations are larger times.

\subsection{Relationship with the corotation torque}
\label{sec:relat-with-hors}
Although the perturbations of density that give rise to the thermal
torques are located in the coorbital region, the thermal torques are
not corotation torques. Corotation torques correspond to the exchange
of angular momentum at a corotation resonance between the perturber
and the disc. The equation that governs heat diffusion is parabolic,
its solution is not wave-like, and a resonant behaviour is impossible.

The derivation presented here is based on a linear decomposition of
the flow's perturbation, and applies in the limit of small planetary
mass. Even in this limit, the flow may eventually exhibit
non-linearities, with a potentially large impact on the torque:
\citet{2009arXiv0901.2265P} have found that, in a disc with a
sufficiently small viscosity, the corotation torque changes its value
over a long time-scale: it adopts initially, over a dynamical
time-scale, the value given by linear theory, then switches to a
different value, given by the (non-linear) horseshoe drag, after a
time-scale which roughly corresponds to the duration of a horseshoe
U-turn\footnote{Over even longer time-scales, the corotation torque
  may exhibit an oscillatory behaviour, and it eventually converges to
  an asymptotic value.}. The smaller the planet mass, the longer it
takes to perform a horseshoe U-turn and therefore to reach the
non-linear regime, but this regime is eventually attained regardless
of the planet mass.

Although the flow may eventually become non-linear in the coorbital
region of a low-mass planet in a disc with thermal diffusion, the
thermal torques should be largely insensitive to this effect, which
should only affect the corotation torque. One may compare the
time-scale of the U-turn \citep{bm08,2009arXiv0901.2265P}
\begin{equation}
  \label{eq:147}
  \tau_\mathrm{U-turn}\sim 10\Omega_p^{-1}\left(\frac{q}{h^3}\right)^{-1/2},
\end{equation}
to the response time of the thermal torques:
$\tau_\mathrm{th}\sim\lambda_c^2/\chi \sim \Omega_p^{-1}$.  For the
largely sub-thermal planets considered in this work, the last factor
in Eq.~\eqref{eq:147} may be substantial, and we expect to have
typically $\tau_\mathrm{U-turn} \sim 10^2\Omega_p^2$, two orders of
magnitude larger than the time it takes to establish the thermal
torque. The horseshoe motion can thus be considered frozen over the
short time-scale required for the heat released at a given instant by
the planet to yield a thermal torque, which should therefore be
insensitive to the horseshoe dynamics. This can also be expected on
intuitive grounds: the perturbation of velocity in the coorbital
region is much smaller than the unperturbed velocity, which is
primarily responsible for the distortion of the hot region that gives
rise to the thermal torque.

\section{Conclusions}
\label{sec:conclusion}

We find that a finite thermal diffusivity changes significantly the
torque experienced by a low-mass planet embedded in a gaseous
protoplanetary disc. We provide an expression, Eq.~\eqref{eq:136}, for
the difference between the torque experienced in a disc with thermal
diffusion and an adiabatic disc, when the planet does not release heat
into the disc. We call this new torque component the cold thermal
torque. It arises from two low-temperature, dense lobes on each side
of corotation. Each one exerts on the planet a torque comparable to
the so-called one-sided Lindblad torque exerted by the outer and inner
legs of the pressure-supported wake. Much like this wake's torque, the
outer disturbance exerts a negative torque, while the inner one exerts
a positive torque. The residual torque, however, is markedly
different. While the relative imbalance of the wakes' torques is of
the order of the disc's aspect ratio $h$, the imbalance of the thermal
torques is set by the ratio of the offset to corotation to the size of
the lobes $\lambda_c\sim\sqrt{\chi/\Omega_p}$. This ratio is usually a
quantity larger than $h$, which implies that the cold thermal torque
is the dominant component of the torque, at least for sufficiently
small planetary masses. We find, by comparing our analytic
expectations to the only numerical evidence of the cold thermal
torques published so far \citep{2014MNRAS.440..683L}, that the cold
thermal torque measured in the simulations has an absolute value
smaller than that of the analytic estimate, even more so as the
planetary mass increases. This suggests that the cold thermal torque
is cut-off for planetary masses above the Earth's mass, and that full
fledged thermal torques have not yet been obtained in numerical
simulations.

When the planet is luminous, the heat released in the surrounding
nebula obeys an equation of advection and diffusion, in which the
advection stems from the Keplerian shear. The relative perturbations
of density and temperature associated with the heat release have
opposite values, while the pressure remains essentially
unperturbed. The perturbation of density associated with the heat
release, in steady state, is asymmetric when the planet is offset from
corotation. We work out an expression for the net torque corresponding
to this perturbation, or heating torque, given by
Eq.~\eqref{eq:144}. This expression is exact in the limit of an offset
to corotation small compared to the size of the disturbance, and of a
size of disturbance small compared to the thickness of the disc. The
total torque is the given by Eq.~\eqref{eq:143}.  As for the cold
thermal torque, the comparison of our analytic expectations to the
values of the heating torque reported by \citet{2015Natur.520...63B}
suggests a cut-off above a fraction of an Earth mass.

Our analysis has assumed the planet to be on a fixed circular
orbit. When the planet's luminosity is large, it can acquire a
significant eccentricity, and the time averaged total torque exerted
on the planet may depart significantly from the circular estimate
\citep{2017arXiv170401931E}.

The decay of the thermal torques below their analytic value for
planetary masses above typically one Earth-mass is not included in the
analytic formulation presented here, based on linear perturbation
theory.  Studying the regime of larger planetary masses requires to
deal with non-linear flows, and will likely require to be tackled
through numerical simulations. This now appears as the most important
step towards an integration of the thermal torques into a general
torque formula.

\section*{Acknowledgements}
The author acknowledges UNAM's PAPIIT grant 101616.




\bibliographystyle{mnras}

\appendix
\section{Solution of the differential systems}
\label{sec:solut-diff-syst}
We seek finite solutions of Eqs.~\eqref{eq:75}-\eqref{eq:76} or
Eqs.~\eqref{eq:94}-\eqref{eq:95}. These solutions are such that the
real part has the parity of the forcing function [$\delta(X)$ in the
first case, $\delta'(X)$ in the second one], whereas the opposite
holds for the imaginary part. Hence, $X\mapsto R_K(X)$ and
$X\mapsto i_K(X)$ are even functions of $X$, while $X\mapsto I_K(X)$ and
$X\mapsto r_K(X)$ are odd functions of $X$.

Integrating Eq.~\eqref{eq:75} from $-\epsilon$ to
$+\epsilon$ and taking the limit $\epsilon\rightarrow 0$, we find
that
\begin{equation}
  \label{eq:148}
  R'_K(0^+)-R'_K(0^-)=2R'_K(0^+)=-1.
\end{equation}
A similar integration of Eq.~\eqref{eq:76} yields that $I'_K$ is
continuous in $0$. Since this function is even in $X$, this implies
that $I''_K(0)=0$ [$I''_K$ cannot be singular in $0$, as per
Eq.~\eqref{eq:76}], and therefore that $I_K(0)=0$.

Similar considerations apply to Eqs.~\eqref{eq:94} and~\eqref{eq:95}.
Integration of Eq.~\eqref{eq:95} with the requisite that $i_K$ is even
implies that $i_K'(0)=0$. Similarly, integration of Eq.~\eqref{eq:94}
on a neighbourhood of $0$ yields
\begin{equation}
  \label{eq:149}
  r_K(\epsilon)\rightarrow
  \frac{\mathrm{sgn}(\epsilon)}{2K}\mbox{~~~for $\epsilon\rightarrow 0$}
\end{equation}

We start our integration at a large value $X_\infty$ of $X$, and integrate
backwards. We consider the two initial conditions
$[R_K(X_\infty),I_K(X_\infty)]=(1,\pm 1)$. We adopt for the first
derivatives the approximate values
\begin{eqnarray}
  \label{eq:150}
  R_K'(X_\infty)&=&\sqrt{\frac{X_\infty}{2K}}[R_K(X_\infty)-I_K(X_\infty)]\\
\label{eq:151}
  I_K'(X_\infty)&=&\sqrt{\frac{X_\infty}{2K}}[R_K(X_\infty)+I_K(X_\infty)].
\end{eqnarray}
We then adopt the linear combination of our two solutions that
verifies $R_K'(0)=-1/(2K)$ and $I_K(0)=0$, and we check in $X_\infty$
that the solution is vanishingly small compared to its value at the
origin. Namely we check that
$|R_K(X_\infty)+iI_K(X_\infty)| < 10^{-9}|R_K(0)+iI_K(0)|$, and
increase $X_\infty$ and repeat the operation until this condition is
satisfied. We have found that our solution, except in the vicinity of
$X_\infty$ where it is vanishingly small, is largely insensitive to
our choice of first derivatives in $X_\infty$. We have found that a
suitable 
choice for $X_\infty$ is
\begin{eqnarray}
  \label{eq:152}
  X_\infty&=&25\mbox{~~~if $K>3$}\\
  &=&17K^{1/3}\mbox{~~~otherwise}
\end{eqnarray}

We use a similar method to solve Eqs.~\eqref{eq:94}-\eqref{eq:95},
which differ only by the forcing term in $X=0$, except that we adopt
the linear combination of our solutions that verifies
$r_K(0^+)=1/(2K)$ and $i'_K(0)=0$.

\section{Numerical evaluation of the integrals of $F$ and $J$}
\label{sec:numer-eval-integr}

The functions $K\mapsto F(K)$ and $K\mapsto J(K)$ are defined respectively
by the integrals that appear in Eqs.~\eqref{eq:81}
and~\eqref{eq:99}. These two functions admit simple approximations in
the limit $K\rightarrow 0$ and $K\rightarrow\infty$. We therefore perform the
calculation of the integrals as follows:
\begin{itemize}
\item We integrate over an interval $[K_\mathrm{min},K_\mathrm{max}]$,
  with $K_\mathrm{min} \ll 1$ and $K_\mathrm{max}\gg 1$, where the
  value of $I_K$ or $i_K$ is obtained using the method described in
  Appendix~\ref{sec:solut-diff-syst}.
\item We add the contribution of the intervals $[0,K_\mathrm{min}]$
  and $[K_\mathrm{max},+\infty]$ in an analytical manner, using the
  approximations respectively at small and large $K$.
\end{itemize}

In the limit $K\rightarrow \infty$, we have the approximation
\begin{equation}
\label{eq:153}
I_K(X)\approx\frac{X+X^2}{8K^2}\exp(-X),
\end{equation}
so that
\begin{equation}
  \label{eq:154}
  F(K)\stackrel{K\rightarrow \infty}{\longrightarrow}\frac{1}{16K^2}.
\end{equation}
In the limit $K\rightarrow 0$, the solution $R_K(X)+iI_K(X)$ tends to
the solution $z_1$ of the equation
\begin{equation}
  \label{eq:155}
  -iXz=Kz''+\delta(X).
\end{equation}
The solution $z_1$ can itself be written in terms of the solution
$z_0$ of the equation
\begin{equation}
  \label{eq:156}
  -ixz=\frac{d^2z}{dx^2}+\delta(x),
\end{equation}
as:
\begin{equation}
  \label{eq:157}
  z_1(X)= K^{-2/3}z_0(XK^{-1/3}),  
\end{equation} 
In the limit of $K\rightarrow 0$ the integral will therefore tend to
\begin{eqnarray}
  \label{eq:158}
  F(K)&\stackrel{K\rightarrow
    0}{\longrightarrow}&\int_{X>0}\Im[Z_K(X)]dX\\
\label{eq:159}
&=&K^{-1/3}\int_{x>0}\Im[z_0(x)]dx
\end{eqnarray}
We find, using a shooting method similar to that of Appendix~\ref{sec:solut-diff-syst}:
\begin{equation}
  \label{eq:160}
  \int_{x>0}\Im[z_0(x)]dx\approx 0.372  
\end{equation}

For the evaluation of $J(K)$, we use
\begin{equation}
  \label{eq:161}
  i_K(X) \stackrel{K\rightarrow
    \infty}{\longrightarrow}\frac{1+X+X^2}{8K^2}\exp(-X),
\end{equation}
so that
\begin{equation}
  \label{eq:162}
  J(K) \stackrel{K\rightarrow
    \infty}{\longrightarrow}\frac{1}{8K^2}
\end{equation}
and, for $K\ll 1$,
\begin{equation}
  \label{eq:163}
  J(K)\approx\frac{0.469}{K^{2/3}}. 
\end{equation}
We obtain:
\begin{eqnarray}
  \label{eq:164}
  \int_0^\infty F(K)dK&\approx& 0.205\\   
  \label{eq:166}
  \int_0^\infty F(K^{2/3})dK&\approx& 0.252\\ 
  \label{eq:167}
  \int_0^\infty J(K^{2/3})dK&\approx& 0.616   
\end{eqnarray}
\section{Enthalpy near the planet}
\label{sec:enthalpy-near-planet}
The derivation of the cold thermal torque requires the knowledge of
the enthalpy distribution in the planet's vicinity in the adiabatic
case. We have made the approximation that the enthalpy is the negative
of the planetary potential. We assess here the degree of accuracy of
this simplifying assumption. There is an indirect albeit simple manner
to evaluate the residual value of the potential plus enthalpy near the
planet. The gas parcels are subjected to an effective potential that
is the sum $\Psi$ of the planetary potential and gas enthalpy.  It is
therefore the depth of this effective potential well that determines
the width of the horseshoe region. Should the enthalpy be exactly the
negative of the potential, the effective potential would vanish and
the horseshoe region would not exist. On the other hand, the fact that
the horseshoe region is much more narrow than what it would be if the
gas parcel were subjected only to the planetary potential (in which
case the horseshoe region would have the same width as in the
restricted three-body problem) indicates that the effective potential
well is much more shallow than the gravitational potential well; that
is, the enthalpy is approximately the opposite of the potential. The
width of the horseshoe region in an adiabatic situation gives us an
idea of the residual value of the effective potential. The Bernoulli
constant on the separatrix of the horseshoe region is
$(3/8)\Omega_p^2x_s^2$, where $x_s$ is the half-width of the horseshoe
region, and it is also the value of the effective potential at the
stagnation point, in the vicinity of the planet \citep{mak2006}. The
effective potential well does not diverge in the vicinity of the
planet and is nearly constant over the innermost pressure
length-scale. We compare it to the planetary potential at the Bondi
radius (which is $c_s^2$, by definition). We have
\begin{equation}
  \label{eq:168}
  \frac{\Psi}{c_s^2}=\frac{3}{8}\frac{\Omega_p^2x_s^2}{c_s^2}=O(\mu/h^3),
\end{equation}
where we have used $x_s\sim r_p(q/h)^{1/2}$. This shows that for
largely sub-thermal planets, our approximation is reasonably
accurate. For an Earth-mass planet in a disc with $h=0.05$, the
assumption that the enthalpy is the negative of the potential well is
accurate to within $\sim 1$~\% at the Bondi radius.

\section{On the heat source of the non-luminous case}
\label{sec:heat-source-non}
The Laplacian of the field $p_a'$ is, using Eq.~\eqref{eq:130}
\begin{equation}
  \label{eq:170}
  \Delta p_a'=-2\nabla\rho_o\cdot\nabla\Phi_p-\rho_0\Delta\Phi_p,    
\end{equation}
where we neglect the second-order derivative of $\rho_0$. We justify
here why we can neglect the first term of the right-hand side of
Eq.~\eqref{eq:170}. The field $\nabla\rho_0$ is uniform on the
neighbourhood of the planet, while the field $\nabla\Phi_p$ is directed
towards the planet. Their dot product is a non-singular field that
changes sign across the planet's orbit. We can estimate the integrated
heat source term $L_+$ of the outer part ($x>x_p$) over a volume
$\lambda_c^3$ as
\begin{equation}
  \label{eq:172}
  |L_+|\sim\frac{\chi|\xi|}{\gamma}\cdot\frac{\rho_0}{r_p}\cdot\frac{GM}{\lambda_c^2}\cdot\lambda_c^3,
\end{equation}
where $\xi=-d\log\rho_0/d\log r$, while the inner part ($x<x_p$) has a
luminosity $L_-$ similar in absolute value, and of opposite sign, to
$L_+$. The one-sided torques excited in the outer disc by $L_+$ and in
the inner disc by $L_-$ have same sign, so the net torque $\Gamma_\pm$
arising from the term of interest is approximately twice the one-sided
torque due to $L_+$, which gives
\begin{equation}
  \label{eq:173}
  \Gamma_\pm \sim 0.13(\gamma-1) \frac{G^2M^2\xi\rho_0}{c_s^2}\cdot\frac{\lambda_c}{r_p}.
\end{equation}
This is approximately $\lambda_c^2/(x_pr_p)$ times the net, cold thermal
torque found in section~\ref{sec:cold-planet}. This ratio is also
$\sim(\lambda_c/H)^2\ll1$, so the torque arising from the term under
consideration is negligible, which justifies our approximation.

\bsp
\label{lastpage}
\end{document}